\documentclass[fleqn,usenatbib]{mnras}

\usepackage[T1]{fontenc}

\DeclareRobustCommand{\VAN}[3]{#2}
\let\VANthebibliography\thebibliography
\def\thebibliography{\DeclareRobustCommand{\VAN}[3]{##3}\VANthebibliography}

\usepackage{graphicx}	
\usepackage{amsmath}	
\usepackage{amssymb}	
\usepackage{siunitx}    
\usepackage{pifont}
\usepackage{newtxtext,newtxmath}

\newcommand{\nut}{{\sc nut}}
\newcommand{\ramses}{{\sc ramses}}
\newcommand{\xmark}{\ding{55}}%
\newcommand{\msun}{\text{M}_{\odot}}
\newcommand{\rgal}{$0.2\,R_{\rm vir, DM}$}

\DeclareSIUnit\erg{erg}

\title[CR feedback in a MW-like galaxy]{The impact of cosmic rays on the interstellar medium and galactic outflows of Milky Way analogues} 

\author[F. Rodr\'iguez Montero et al.]{Francisco Rodr\'iguez Montero,$^{1}$\thanks{E-mail: currodri@gmail.com}
Sergio Martin-Alvarez,$^{2}$
Adrianne Slyz,$^{1}$
Julien Devriendt,$^{1,3}$
\newauthor
Yohan Dubois,$^{4}$ and Debora Sijacki$^{5}$ 
\\
$^{1}$Sub-department of Astrophysics, University of Oxford, Keble Road, Oxford OX1 3RH, UK\\
$^{2}$Kavli Institute for Particle Astrophysics and Cosmology (KIPAC), Stanford University, Stanford, CA 94305, USA\\
$^{3}$CNRS, Centre de Recherche Astrophysique de Lyon, Universit\'e de Lyon, Universit\'e Lyon 1, ENS de Lyon, UMR 5574, F-69230 Saint-Genis-Laval, France\\
$^{4}$Institut d’Astrophysique de Paris, UMR 7095, Sorbonne Universit\'e, CNRS, 98 bis boulevard Arago, 75014 Paris, France\\
$^{5}$Institute of Astronomy and Kavli Institute for Cosmology, University of Cambridge, Madingley Road, Cambridge CB3 0HA, UK\\
}

\date{MNRAS, submitted}

\pubyear{2023}

\begin{document}
\label{firstpage}
\pagerange{\pageref{firstpage}--\pageref{lastpage}}
\maketitle

\begin{abstract}
During the last decade, cosmological simulations have managed to reproduce realistic and morphologically diverse galaxies, spanning the Hubble sequence. Central to this success was a phenomenological calibration of the few included feedback processes, whilst glossing over higher complexity baryonic physics. This approach diminishes the predictive power of such simulations, preventing to further our understanding of galaxy formation. To tackle this fundamental issue, we investigate the impact of cosmic rays (CRs) and magnetic fields on the interstellar medium (ISM) and the launching of outflows in a cosmological zoom-in simulation of a Milky Way-like galaxy. We find that including CRs decreases the stellar mass of the galaxy by a factor of 10 at high redshift and $\sim 4$ at cosmic noon, leading to a stellar mass to halo mass ratio in good agreement with abundance matching models. Such decrease is caused by two effects: i) a reduction of cold, high-density, star-forming gas, and ii) a larger fraction of SN events exploding at lower densities, where they have a higher impact. SN-injected CRs produce enhanced, multi-phase galactic outflows, which are accelerated by CR pressure gradients in the circumgalactic medium of the galaxy. While the mass budget of these outflows is dominated by the warm ionised gas, warm neutral and cold gas phases contribute significantly at high redshifts. Importantly, our work shows that future JWST observations of galaxies and their multi-phase outflows across cosmic time have the ability to constrain the role of CRs in regulating star formation.

\end{abstract}




\section{Introduction}\label{sec:introduction}
Realistically modelling galaxy formation within the $\Lambda$CDM cosmological paradigm has proven to be an extremely complex task, as its multi-scale and multi-physics nature quickly develops a non-linear behaviour that is difficult to grasp and model. The most sophisticated amongst modern cosmological simulations have confronted this challenge by explicitly solving the equations governing the evolution of matter across the vast range of scales spanning from the cosmic large-scale structure down to the interstellar medium (ISM) of individual galaxies. For processes such as star formation which are unresolved, yet indispensable for galaxy formation, numerical prescriptions are implemented. Much of their success \citep[for a review see e.g.][and references therein]{Vogelsberger2020} resides in their recognition that, without the injection of energy by clustered supernova (SN) events, the gravitational collapse of cold-dense gas leads to the consumption of the gas reservoir over a free-fall time, resulting in too-efficient star formation \citep{Kay2002,Bournaud2010ISMFeedback,Dobbs2011,Hopkins2011,Tasker2011,Silk2010FeedbackFormation,Moster2013GalacticHaloes} in stark difference with observations \citep[e.g.][]{Zuckerman1974ModelsClouds,Williams1997TheClouds,Kennicutt1998,EvansII1999PhysicalFormation,Krumholz2007SlowImplications,Evans2009TheLifetimes}. Besides the need for reproducing the inefficient formation of stars \citep[e.g.][]{Stinson2006,Teyssier2013,Springel2003}, including such a considerable energy injection through SN or active galactic nuclei (AGN) establishes a feedback loop which is of utter importance for reproducing many global observables of the galaxy population. Illustrative examples are the mass-metallicity relation \citep{Tremonti2004TheSDSS,Keres2009GalaxiesFeedback,Mannucci2010AGalaxies,Maiolino2019DeGalaxies,Tortora2022ScalingFactors}, the metal enrichment of the circum- (CGM) and intergalactic medium (IGM) \citep[e.g.][]{Veilleux2005,Brooks2007TheSimulations,Chisholm2018Metal-enrichedRelationship}, the quenching and morphological transformation of late-type galaxies into early-type galaxies \citep[e.g.][]{DeLucia2007TheGalaxies,Naab2007FormationConditions,Dubois2013AGN-drivenGalaxies,Dubois2016TheFeedback}, the flattening of inner dark matter profiles \citep[e.g.][]{Mashchenko2006TheUniverse,Governato2010BulgelessOutflows,DeBlok2010TheProblem,Teyssier2013} and the formation of bulge-less galaxies by expelling low angular momentum gas \citep{Governato2012CuspyGalaxies}.

However, this success is built on a phenomenological strategy, in which subgrid models for feedback processes are calibrated against local observed scaling relations \citep[e.g.][]{Crain2015TheVariations,Pillepich2018, Rosdahl2018,Dave2019} thus considerably curtailing predictive power. For stellar feedback, state-of-the-art simulations are now moving into a regime where the multi-phase structure of the interstellar medium is beginning to be spatially resolved. This has driven the development of more physically motivated treatments of SNe feedback \citep[e.g.][]{Hopkins2014,Kimm2014EscapeStars,Kimm2015}, as well as other forms of stellar feedback that could not be accounted for in the previous generations of numerical simulations such as e.g. stellar radiation, including photoionisation heating and radiation pressure \citep{Murray2005,Hopkins2011,Agertz2013TOWARDSIMULATIONS,Rosdahl2015ARAMSES-RT,Emerick2018StellarGalaxies,Smith2018SupernovaNumerics,Smith2019CosmologicalAlone}, stellar winds \citep{Hopkins2011,Agertz2013TOWARDSIMULATIONS,Geen2015AStar,Fierlinger2016StellarWinds,Smith2019CosmologicalAlone}, or direct pressure from Lyman-$\alpha$ photons \citep{Kimm2018}. Although simulations are beginning to resolve the scales and physical processes required to model star formation through prescriptions aimed to reproduce their formation in giant molecular clouds, almost all aspects of the actual launching of outflows at galactic scales and their thermodynamic properties remains a source of intense debate in the theoretical and numerical simulations community. Furthermore, the observations that are used to constrain stellar feedback models \citep[e.g.][]{Collins2022ObservationalGalaxies,Leroy2023PHANGSJWSTEra,Thilker2023PHANGSJWSTActivity} suffer from large limitations, with reported mass-loading factors (defined as the ratio of gas mass ejected from the galaxy to the current rate of gas depletion by star formation) varying by more than 2 orders of magnitude in galaxies of similar mass \citep[e.g.][]{Veilleux2005,Chisholm2016A6090}. There are also considerable difficulties involved in detecting the emission from hot ($\sim 10^6$ K) gas \citep{Heckman1995AnOutflow,Summers2003ChandraGalaxy,McQuinn2019GalacticGalaxies,Marasco2023ShakenGalaxies,Concas2022BeingHosts}, which theoretical models argue carry a significant fraction of the kinetic energy of the wind \citep[e.g.][]{Kim2018NumericalModel}.

Cosmic rays (CRs), relativistic particles which are accelerated by diffusive shock acceleration \citep[e.g.][]{Axford1981THEWAVES,Bell1978a,Blandford1987ParticleOrigin}, act as a non-thermal source of energy, and pioneering numerical investigations \citep[e.g.]{Jubelgas2008,Wadepuhl2011,Booth2013,Salem2014} of their impact indicate that they can have a significant impact on star formation. Observations of the Milky Way and nearby galaxies suggest that CRs and magnetic fields are in equipartition with the thermal and turbulent energies in their mid-plane \citep{Boulares1990GalacticDiffusion,Basu2013MagneticEquipartition,Beck2016MagneticGalaxies,Zweibel2017,Lopez-Rodriguez2021The82}. Furthermore, the peak of their spectrum at $\sim 1$ GeV leads them to exhibit smooth spatial gradients on $\gtrsim 1$ kpc scale lengths \citep[e.g.][]{Ennslin2004}. Consequently, for some time now they have been put forward as a fundamental player in the dynamics of extra-planar gas and the launching of outflows \citep[e.g.][]{Ipavich1975,Breitschwerdt1991GalacticGalaxy.,Ptuskin1997TransportRays.,Socrates2006TheGalaxies,Everett2008,Mao2018GalacticPressure,Quataert2021TheDiffusion,Quataert2021TheSolutions}. These $\sim 1$ GeV CRs behave as a relativistic fluid with a soft equation of state (i.e. adiabatic index $\gamma_{\rm CR}=4/3$), thus making the mixture of CRs and non-relativistic gas more compressible and undergoing smaller energetic losses under adiabatic expansion. Additionally, for average low-redshift ISM conditions, CRs experience longer cooling times ($\sim 10^{6-7}$ Myr) via Coulomb and hadronic interactions \citep{Ennslin2004,Schlickeiser2009Non-linearElectrons} than the radiative cooling of thermal gas. They rapidly diffuse away from their injection sites due to diffusion lengths larger than typical molecular clouds ($\sim 100$ pc), reaching the more dilute gas immediately above the galactic disk, which is easier to accelerate against the gravitational potential. Therefore, in the past few years there has been an increased effort in modelling CR feedback in numerical studies of galactic outflows, showing broadly consistent trends across hydrodynamic schemes and numerical implementations of the CR Fokker-Planck equation \citep[][and references therein]{Skilling1971CosmicDiffusion,Hanasz2021SimulationsPropagation}: (i) CRs thicken the gaseous disk by altering the pressure-gravity balance, changing the morphology of the galaxy \citep[][]{Booth2013,Salem2014,Ruszkowski2017,Dashyan2020,Chan2022TheGalaxies,Farcy2022Radiation-MagnetoHydrodynamicsGalaxies,Nunez-Castineyra2022Cosmic-rayLuminosities,Buck2020TheContext,Martin-Alvarez2022TheGalaxies}. In addition, CR feedback significantly affects the star formation history of galaxies by (ii) reducing star formation rates \citep[e.g.][]{Jubelgas2008,Booth2013,Pakmor2016,Chan2019,Dashyan2020,Hopkins2020a} and by (iii) launching more heavily mass-loaded outflows, these being denser and colder \citep[e.g.][]{Booth2013,Hanasz2013,Hopkins2021CosmicLgalaxies,Farcy2022Radiation-MagnetoHydrodynamicsGalaxies,Peschken2022TheGalaxies}.

This modelling progress has established that CRs in the $\sim 1$ GeV regime are likely an important ingredient of galaxy formation. However, a clear understanding of its coupling with other physical processes of relevance remains far from being explored. While isolated galaxy simulations find CRs modelled with a constant diffusion coefficient $\kappa$ are able to affect the star formation of galaxies over a range of masses \citep{Booth2013,Ruszkowski2017,Dashyan2020,Farcy2022Radiation-MagnetoHydrodynamicsGalaxies,Nunez-Castineyra2022Cosmic-rayLuminosities}, their inclusion in cosmological simulations shows \citep[e.g.][]{Jubelgas2008,Buck2020TheContext,Martin-Alvarez2022TheGalaxies} that their effect on the final stellar mass of a galaxy is usually small, preferentially affecting the most massive galaxies at low redshift \citep[e.g.][]{Hopkins2020a}. This is likely due to complex interplay between CR-driven outflows and cosmic gas inflows in a realistic CGM, that can be captured only by fully self-consistent cosmological simulations \citep[see e.g.][]{Ji2020PropertiesHaloes, Ji2021VirialHaloes, Beckmann2022CosmicClusters, Martin-Alvarez2022TheGalaxies}. 
 
Although $\sim 1$ GeV transport is usually modelled in the streaming instability limit (i.e. CR-induced magnetic turbulence of the order of the particle gyroradius), more physically motivated descriptions of CR transport suggest that in the warm and cold phases Alfv\'en waves are efficiently damped by ion and neutral collisions and by MHD turbulence \citep[e.g.][]{Farmer2003WaveISM,Xu2022CosmicMedium}, giving rise to a variation of up to 5 orders of magnitude in the CR mean free path \citep{Armillotta2022Cosmic-RayEnvironments} and to changes in the acceleration of cold clouds in the CGM \citep{Bruggen2020TheAcceleration, Bustard2021Cosmic-RayMedium}. Moreover, CRs can affect the evolution of magneto-thermal instabilities in galactic discs depending on the transport prescription and diffusion constant \citep[e.g.][]{Wagner2005CosmicInstability,Shadmehri2009ThermalDiffusion,Kuwabara2020DynamicsRays}, strongly altering its structure \citep{Commercon2019Cosmic-rayMedium,Dashyan2020,Nunez-Castineyra2022Cosmic-rayLuminosities,Martin-Alvarez2022TheGalaxies} and the conditions in which star formation and stellar feedback take place. Hence, an accurate modelling of the ISM is fundamental for a better understanding of CR feedback together with the launching and acceleration of CR-driven multi-phase outflows \citep[][]{Girichidis2016LaunchingMedium, Kim2018NumericalModel, Salem2014,Hopkins2021CosmicLgalaxies,Martin-Alvarez2022TheGalaxies}, which remains to be carefully studied in a cosmological context.

This is the first paper in a series in which we study the effects of CR feedback in galaxies with a resolved multi-phase ISM using high-resolution cosmological zoom-in  simulations, with the aim to further our understanding of the role CRs play in shaping galaxy properties. In Section~\ref{sec:methods} we outline our Milky Way-like galaxy (called \nut) formation simulation set-up, and our CR and magnetohydrodynamics modelling. We then discuss the global effects of magnetic fields and CRs in the evolution of \nut~in Section~\ref{subsec:global_effects}, demonstrating how the CRs reduce the stellar mass by decreasing the star formation efficiency in Section~\ref{subsec:sf_efficiency} and increasing the SN feedback efficiency in Section~\ref{subsec:sn_outflows} via modifications to the outflow phases and their more efficient acceleration. In Section~\ref{sec:conclusions} we summarise our main results, discuss their limitations and the prospect for future studies of CR feedback. 

\section{Numerical Methods}\label{sec:methods}
\begin{figure*}
	\includegraphics[width=\textwidth]{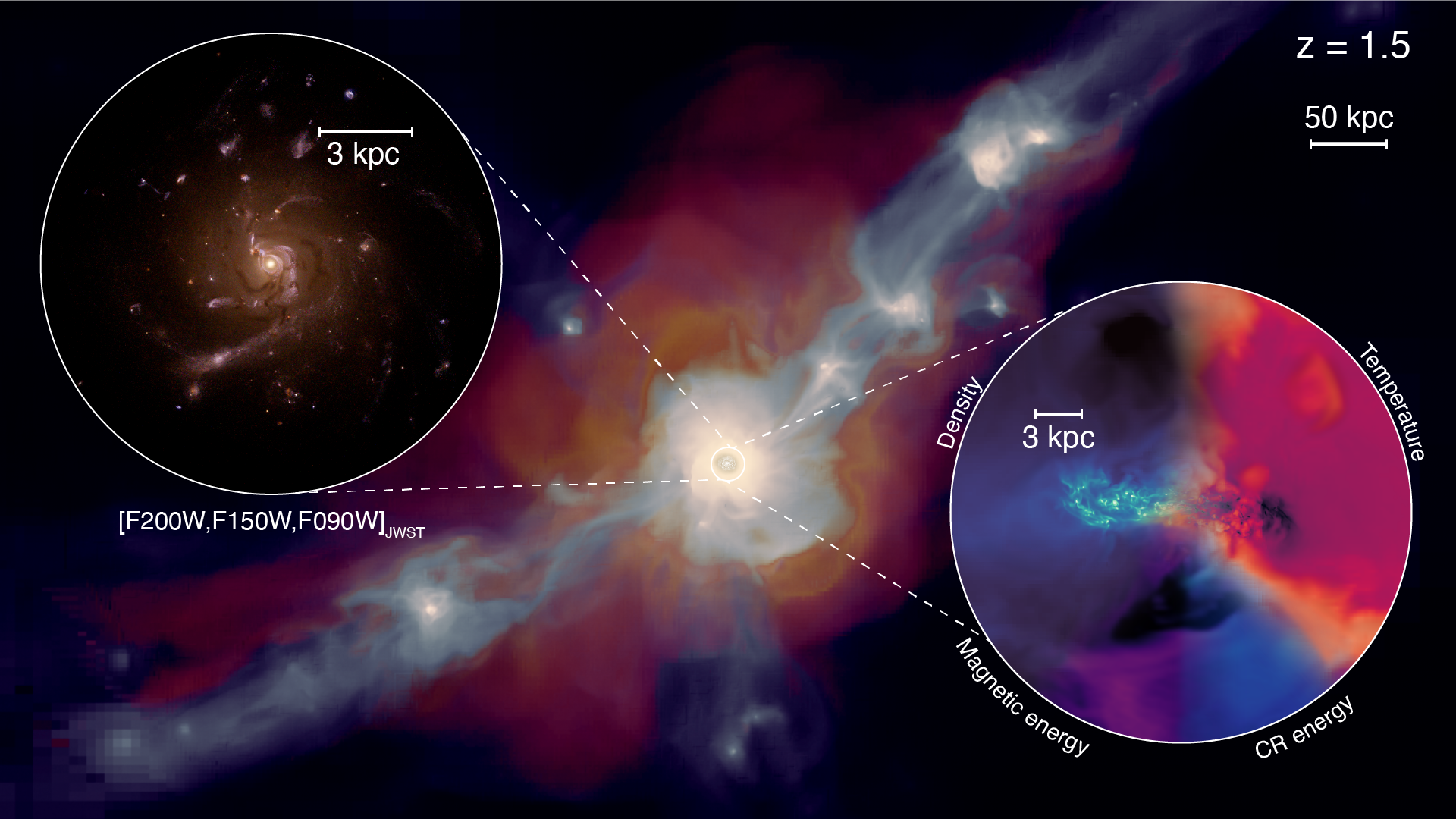}
    \caption{Overview of the CRMHD model of the \nut~simulation at $z=1.5$. \textbf{Background:} Colour composite of the zoom-in region centred on the \nut~galaxy, with gas density (silver blue), gas temperature (orange) and CR energy density (yellow), showing 600~kpc along the x-axis. Three cold filaments are feeding the growth of the central galaxy, where strong feedback events are driving large-scale outflows of hot and CR-dominated gas. The effects of feedback are also observed in smaller objects embedded in the infalling filaments. \textbf{Left inset:} RGB colour composite zoom into the immediate galactic region as would be observed with the F200W, F150W and F090W filters from JWST. This mock observation is generated using the {\sc SKIRT} radiation transfer code by assuming a fixed dust-to-metal ratio of 0.4 and the {\sc BARE-GR-S} dust model from \protect\cite{Zubko2004InterstellarConstraints}. \textbf{Right inset:} Edge-on view of the \nut~galaxy presenting different gas quantities modelled in the CRMHD simulation. In clockwise order from top-left corner: gas density, gas temperature, CR energy density, and magnetic field energy density. The cold and dense disk is shown in this projection, combined with a large hot outflow along the axis of rotation of the galaxy. CR and magnetic energies fill the halo and are shaped by a recent outflows.}
    \label{fig:canvas}
\end{figure*}

\subsection{The RAMSES code}

The full suite of new simulations presented in this work has been generated using the \ramses~code\footnote{\href{https://bitbucket.org/rteyssie/ramses/src/master/}{https://bitbucket.org/rteyssie/ramses/src/master/}} \citep{Teyssier2002}, a grid-based Eulerian code with an octree AMR grid. \ramses~uses collisionless particles to model the evolution of the dark matter and stellar components. These are coupled to each other and to the gas component through gravity described by the Poisson equation solved on the adaptive mesh. In order to evolve in time the magneto-hydrodynamical quantities of the gas, it makes use of a second-order Godunov scheme with the Harten-Lax-van Lear-Discontinuity Riemann solver from \cite{Miyoshi2005AMagnetohydrodynamics}, and the MinMod slope limiter to reconstruct the cell-centred properties. The implementation of magnetic fields in \ramses~uses the constrained transport (CT) method, which models magnetic field values as cell face-centred quantities \citep{Teyssier2006,Fromang2006}. This method ensures that the solenoidal condition (i.e. $\vec{\nabla}\cdot \vec{B} = 0$) is fulfilled to within numerical precision errors, preventing the numerically spurious growth of magnetic field and cell noise \citep[see][for a comparison of CT with other MHD methods]{Toth2000TheCodes,Mocz2016AMagnetohydrodynamics}.

\subsubsection{Cosmic ray magneto-hydrodynamics}
We make use of the implementation of CRs from \cite{Dubois2016} and \cite{Dubois2019}, in which CRs are treated as a separate fluid represented by one additional energy equation in the set of MHD fluid equations:
\label{ap:crmhd_equations}
\begin{align}
    &\frac{\partial \rho}{\partial t} + \vec{\nabla} \cdot (\rho\vec{u}) = 0,\\
    &\frac{\partial \rho \vec{u}}{\partial t} + \vec{\nabla} \cdot \left(\rho \vec{u}\vec{u} + P_{\rm tot}- \frac{\vec{B}\vec{B}}{4\pi} \right) =\rho \vec{g},\\
    &\frac{\partial e}{\partial t} + \vec{\nabla} \cdot \left((e + P)\vec{u} - \frac{\vec{B} (\vec{B}\cdot \vec{u})}{4\pi} \right) \nonumber\\
    &=\rho \vec{u}\cdot \vec{g}-P_{\text{CR}}\vec{\nabla} \cdot \vec{u} - \vec{\nabla} \cdot \vec{F}_{\text{CR,d}}+ \mathcal{L}_{\text{rad}}, \label{eq:e}\\
    &\frac{\partial \vec{B}}{\partial t} - \vec{\nabla} \times (\vec{u}\times\vec{B}) =0,\label{eq:induction}\\
    &\frac{\partial e_{\text{CR}}}{\partial t} + \vec{\nabla} \cdot (e_{\text{CR}}\vec{u} + (e_{\text{CR}} + P_{\text{CR}})\vec{u}_{\text{st}}) \nonumber\\
    &= -P_{\text{CR}}\vec{\nabla} \cdot \vec{u} - \vec{\nabla} \cdot \vec{F}_{\text{CR,d}}+ \mathcal{L}_{\text{st}} + \mathcal{L}_{\text{rad,CR}, \label{eq:e_cr}}
\end{align}
where $\rho$ is the gas mass density, $\vec{u}$ is the gas velocity vector, $\vec{g}$ the gravitational field, $\vec{u}_{\text{st}}=-\vec{u}_{\text{A}}\text{sign}(\vec{b}\cdot \vec{\nabla} e_{\rm CR})$ is the streaming velocity (where $\vec{u}_{\text{A}}$ is the Alfvén velocity and $\vec{b} = \vec{B}/\vert \vec{B}\vert$ the magnetic unit vector), $\vec{B}$ is the magnetic field, $e=0.5\rho \vert\vec{u}\vert^2+e_{\text{th}}+e_{\text{CR}}+\vert \vec{B}\vert^2/(8\pi)$ is the total energy density, $e_{\text{th}}$ is the thermal energy density, and $e_{\text{CR}}$ is the CR energy density, $P_{\rm tot}=P_{\text{th}}+P_{\text{CR}}+P_{\text{mag}}$ is the sum of the thermal $P_{\text{th}}=(\gamma - 1)e_{\text{th}}$, CR $P_{\text{CR}}=(\gamma_{\text{CR}} - 1)e_{\text{CR}}$ and magnetic $P_{\text{mag}}=\vert \vec{B}\vert^2/(8\pi)$ pressures, with adiabatic indexes $\gamma=5/3$ and $\gamma_{\rm CR}=4/3$ for an ideal non-relativistic and relativistic gas, respectively. This modelling approximation can be adopted for the evolution of low-energy ($\sim $GeV) CRs, as their interaction with the thermal gas is mostly collisionless and mediated by pitch-angle scattering by the local magnetic field. A practical way of interpreting the terms on the right side of Equations~(\ref{eq:e}) and (\ref{eq:e_cr}) is to treat them as source terms, with $P_{\text{CR}}\vec{\nabla} \cdot \vec{u}$ the CR pressure work, $\vec{F}_{\rm CR,d}=-\kappa \vec{b}(\vec{b}\cdot \vec{\nabla} e_{\rm CR})$ the anisotropic diffusion flux ($\kappa$ is the diffusion coefficient, assumed constant in this work), $\mathcal{L}_{\rm rad}=\mathcal{L}_{\rm rad,th} + \mathcal{L}_{\rm rad,CR \rightarrow th}$ the total radiative losses which includes thermal and CR radiative losses, respectively\footnote{The modelling of radiative losses is described in Section~\ref{subsec:nut_methods}.}. The CR loss term encompasses both energy losses due to hadronic and Coulomb interactions $\mathcal{L}_{\rm CR}=-7.51\times 10^{-16}(n_{\rm e}/\text{cm}^{-3})(e_{\rm CR}/\text{erg cm}^{-3})\text{ erg s}^{-1}$ \citep[see e.g.,][]{Ennslin2004,Guo2008} transferred to the thermal component at a rate $\mathcal{H}_{\rm CR \rightarrow th}=2.63\times 10^{-16}(n_{\rm e}/\text{cm}^{-3})(e_{\rm CR}/\text{erg cm}^{-3})\text{ erg s}^{-1}$.

In the CR energy equation (Equation~(\ref{eq:e_cr})), $\vec{F}_{\rm st}=(e_{\rm CR}+P_{\rm CR})\,\vec{u}_{\rm st}$ is the streaming advection term, which emulates the self-confinement of CR transport by the self-excited Alfv\'en waves \citep{Kulsrud1969,Skilling1975CosmicParticles}. As we have discussed in Section~\ref{sec:introduction}, several mechanisms may play an important role in damping these self-excited Alfv\'en waves in different gas phases \citep[e.g.][]{Armillotta2022Cosmic-RayEnvironments}, allowing for streaming velocities exceeding $\vec{u}_{\rm A}$. As a preliminary step towards a deeper understanding of CR feedback in a cosmological context, we assume that CR streaming velocities cannot reach super-Alfv\'enic velocities, and leave the implementation of self-consistent CR diffusion coupling \citep[like the one presented in][]{Farber2018ImpactWinds,Xu2022CosmicMedium,Armillotta2022Cosmic-RayEnvironments} for future work. Furthermore, since CRs scatter on the Alfv\'en waves, they experience a drag force, transferring energy to the thermal gas at a rate $\mathcal{L}_{\rm st}=-\text{sign}(\vec{b}\cdot \vec{\nabla} e_{\rm CR}) \vec{u}_{\rm A}\cdot \vec{\nabla} P_{\rm CR}$. The typical value of $\kappa \simeq 3\times 10^{28}(R_{\rm CR}/1\text{GV})^{0.34}\text{cm}^2 \text{ s}^{-1}$ depends on the particle rigidity $R_{\rm CR}$ \citep{Johannesson2019Cosmic-RayGeminga}, which for 1 GeV proton has a value of $R_{\rm CR}=1 \text{ GV}$. Previous efforts to constrain $\kappa$ in CR simulations by comparing Fermi LAT $\gamma$-ray observations\citep[e.g. {\sc Fire}-2, {\sc arepo} and \ramses,][respectively]{Chan2019,Werhahn2021CosmicCorrelation,Nunez-Castineyra2022Cosmic-rayLuminosities} have resulted in large discrepancies in the inferred diffusion coefficient favoured by observations. \cite{Nunez-Castineyra2022Cosmic-rayLuminosities} argue that, although differences in the assumed CR spectrum or $\gamma$-ray emissivity per atom can reduce the discrepancy by factors of up to 2.5, these are not sufficient to bring different simulations to agreement. They suggest that differences in the thermal state of the gas and the magnetic field properties, achieved by different MHD codes, could be the cause of these differences. The exploration of how different CR transport mechanisms and diffusion coefficients affect the $\gamma$-ray luminosity is the scope of an upcoming paper (Rodr\'iguez Montero et al., \textit{in prep}). We adopt here a value of $3\times 10^{28} \text{cm}^2\,\text{s}^{-1}$ favoured by the results of isolated CR simulations using similar versions of \ramses~to the one used in this work \citep{Dashyan2020,Farcy2022Radiation-MagnetoHydrodynamicsGalaxies,Nunez-Castineyra2022Cosmic-rayLuminosities}, and leave the analysis of a higher $\kappa=3\times 10^{29} \text{cm}^2\,\text{s}^{-1}$ \citep[favoured by the {\sc Fire}-2 simulations][]{Chan2019,Hopkins2021TestingEnergies} to future work.

\subsection{The \nut~suite of simulations}\label{subsec:nut_methods}

The new suite of simulations (see Table~\ref{table:nut_suite} for details) presented in this work uses the initial conditions of the \nut~suite (\cite{Powell2011} and \cite{Geen2013SatelliteHaloes}) generated at $z=499$ with cosmological parameters consistent with the results of WMAP5 \citep[matter density $\Omega_{\rm m}=0.258$, dark energy density $\Omega_\Lambda=0.742$, baryon density $\Omega_{\rm b}=0.045$, amplitude of the linear power spectrum at the $8\,\rm Mpc$ scale $\sigma_8=0.8$ and Hubble constant $H_0=72\,\rm km\,s^{-1}\, Mpc^{-1}$][]{Dunkley2009Five-yearData}. The computational box has a side of comoving length 9 Mpc $h^{-1}$ with a spherical region of approximately 2.7 comoving Mpc $h^{-1}$ in diameter defining the volume in which high resolution is performed. Within this zoom-in region a $M_{\rm vir}(z=0)\simeq 5 \times 10^{11} M_\odot$ DM halo hosts a Milky Way-like galaxy (from this point simply identified as the \nut~galaxy) with a DM particle resolution of $m_{\rm DM}=5.5\times 10^4 M_\odot$. To probe the multi-phase structure of the ISM, we allow refinement within this region to be triggered via a quasi-Lagrangian method, for which cells with total mass (i.e. gas, stars and dark matter) above $8 m_{\rm DM}\Omega_{\rm m}/\Omega_{\rm b}$ are subdivided into 8 equal cells, reaching a maximum physical resolution of cell size $\Delta x_{\max}\simeq 23$ pc approximately constant in proper units (an extra level of refinement is triggered for expansion factors of $a_{\rm exp}=0.1,0.2,0.4,0.8$).

In Fig.~\ref{fig:canvas} we show an overview of our fiducial simulation with CRs and MHD centred on \nut. The main galaxy is fed by three large filaments (gas density in silver), and at their intersection there is a gas halo filled with high temperature gas (in orange) and CR energy (yellow) from recent feedback events. Additionally, we show zoom-in insets of the galactic region of \nut~ seen face-on (left circle) as it would be seen through JWST F200W (red), F150W (green) and F090W (blue) filters, and edge-on (right circle) density-weighted projections of the raw simulation data. The mock JWST RGB image has been obtained with the radiation transfer code {\sc skirt} \citep{Camps2015SKIRT:Architecture}, assuming a fixed dust-to-metal ratio of 0.4 and the {\sc BARE-GR-S} dust model from \cite{Zubko2004InterstellarConstraints}. At $z=1.5$ \nut~shows signatures of a large feedback event: hot bipolar outflows emerge perpendicular to the cold, dense and magnetised disk. These outflows clear denser and colder gas in the CGM, creating bubbles seen in the gas density, magnetic energy and CR energy. The face-on view through SDSS filters shows a late-type spiral with a nuclear cluster at its centre dominated by an old stellar population, and a disk filled with lanes and filaments of dense dust absorption and small regions of recent star formation.

For all our runs, we include metal-dependent radiative cooling with interpolated cooling curves from {\sc cloudy} \citep{Ferland1998}, down to $\sim10^4$ K. Below this temperature, gas is allowed to cool further via metal fine structure atomic transitions \citep{Rosen1995} (to a minimum of 15 K) and via adiabatic expansion. We employ a uniform ultraviolet background that is turned on at a redshift of $z=9$ \citep{Haardt1996}, with gas denser than $n_{\rm H}=0.01$ cm$^{-3}$ considered to be self-shielded from UV radiation with an exponential dependence on $n_{\rm H}$. Furthermore, despite the high resolution of our simulation, we lack the cooling modelling capabilities to resolve primordial (Population III) SN explosions in minihalos \citep[e.g.][]{Whalen2008TheSupernovae,Wise2012ThePopulations} that would potentially raise the initial gas metallicity to $\sim 10^{-3}\,Z_\odot$ and hence increase cooling at high redshift. To avoid this issue, we initialise the simulation with a uniform metallicity floor of $10^{-3}\,Z_\odot$.

The formation of stellar particles from the gas only takes place in the cells at the highest level of refinement when the gas is found to be gravitationally unstable. We determine this following the magneto-thermo-turbulent (MTT) star formation model presented in \cite{Kimm2017Feedback-regulatedReionisation} and in \cite{Martin-Alvarez2020} for its MHD version. Star formation can only happen where the gas number density is above 10 cm$^{-3}$ and the gravitational pull is larger than the combined support from turbulent, thermal, magnetic and CR pressure\footnote{We have checked the change in efficiency due to the addition of CR pressure to the original MTT model, and found that it commonly decreases the efficiency by $\sim 10-30$\%, although in rare occasions the support of CR pressure can be enough to turn the efficiency to 0.}. We determine the rate of gas to stellar mass conversion following a Schmidt law \citep{Schmidt1959}
\begin{align}
    \dot{\rho}_{\rm star} = \epsilon_{\rm ff}\frac{\rho_{\rm gas}}{t_{\rm ff}},
\end{align}
where we allow $\epsilon_{\rm ff}$ to be a local parameter that is determined by the gas cell conditions as predicted by the multi-free-fall model from \cite{Hennebelle2011AnalyticalFragmentation} (generalised to the model of PN11 by \cite{Federrath2012TheObservations}), and refer the reader interested in further details of this model to appendix B in \cite{Martin-Alvarez2020}. Ultimately, the conversion of gas to stars is modelled by sampling stochastically a Poisson mass-probability distribution for the final stellar particle mass~\citep{Rasera2006TheBudget}, with the minimum stellar mass at creation fixed at $4.5\times 10^3M_\odot$.

As discussed in the introduction, instead of performing a calibration process to determine artificial `boosting' factors to the canonical type II SN energy of $10^{51}$~erg that may allow, e.g., the recovery of a stellar-to-halo-mass (SMHM) relation in agreement with local observations \citep{Rosdahl2018}, we model SN feedback using the `mechanical' approach of \cite{Kimm2014EscapeStars}. Considering the local gas conditions and the available spatial resolution, it determines whether the Sedov-Taylor expansion phase of the SN can be captured directly \citep{Blondin1998,Thornton1998} or if numerical overcooling will artificially affect its evolution. In the latter case, momentum prescribed by the snowplow phase is injected. In order to determine when a stellar particle should experience a SN feedback event, we follow the \textit{multiple explosions} method by \citet{Kimm2015}, in which we use the real delay of SNe given by the population synthesis code {\sc starburst99} \citep{Leitherer1999Starburst99:Formation}, with SNe taking place as early as 3~Myr and as late as 50~Myr. These SN events inject mass, momentum and energy into the host and a maximum number of 48 neighbouring cells, with a canonical efficiency of $\varepsilon_{\rm SN}=10^{51}\text{ erg}/(10M_\odot)$ and a fraction of 0.2 of the SN mass returned to the SN host cell. Additionally, a fraction 0.075 of this total mass corresponds to metals, which are followed via a passive scalar advected with the gas flow by the Godunov solver. We further assume a Kroupa IMF \citep{Kroupa2001} for the underlying stellar population of our star particles. 

The ideal MHD induction equation prohibits the formation of magnetic fields in the absence of an initial non-zero $\vec{B}$ field. Several mechanisms have been presented over the last decades as candidates to generate galactic magnetic fields of the order of $10^{-7}-10^{-5}$~G \citep[e.g.][]{Mulcahy2014,Beck2016MagneticGalaxies}. These can be classified by their cosmological \citep[e.g., electroweak or quantum chromo-dynamics phase transitions][]{Vachaspati1991MagneticTransitions,Olesen1993AString} or astrophysical \citep[plasma seeding processes like Biermann battery][]{Biermann1950UberSchluter,Pudritz1989TheProtogalaxies} nature \citep[see][for recent reviews on the matter]{Durrer2013CosmologicalObservation,Vachaspati2021ProgressFields}. Whether magnetic fields in galaxies are primordial in nature or rapidly amplified by astrophysical or dynamo processes \citep{Pakmor2013SimulationsGalaxies,Rieder2017,Vazza2017TurbulenceFormation,Martin-Alvarez2018,Martin-Alvarez2021}, the theoretical expectation is that they reach $\sim 10^{-6}$~G strengths no later than a few hundred Myr after galaxy formation \citep{Martin-Alvarez2022TheGalaxies}. We permeate the simulation box with an uniform magnetic field along the $z$ axis\footnote{In this work we do not explore the effect of different orientations, as this does not significantly alter the evolution of \nut~\citep{Martin-Alvarez2020}, but the magnetisation of the large scale environment changes.}. In order to mimic such rapid amplification, we employ a uniform primordial magnetic field of strength $B_0=3 \times 10^{-12}$~G. This simple magnetic field initialisation directly provides the expected strengths of~$1-10$ $\mu$~G in the \nut~galaxy shortly after its formation \citep{Martin-Alvarez2021}. 

In order to properly capture CR evolution, we activate the CR solver when the first stars are formed: shortly prior to their first injection into the simulation. This is because SNe are the exclusive source of CRs in our simulations. We assume that each SN event injects a fixed fraction of its total energy in the form of CR energy into the gas cell hosting the SN particle. We choose this fraction to be $f_{\rm CR}=0.1$, which leads to a CR injected energy per SN event of $10^{50}$ erg. This value is in fair agreement with the estimated acceleration efficiency of CRs by non-relativistic shocks \citep[e.g.][]{Morlino2012,Caprioli2014}. In order to keep the energy injection for type II SN fixed to the canonical value, the energy available for thermal energy and radial momentum injection by the mechanical feedback prescription is reduced accordingly.

Simulation snapshots are analysed using {\sc Ozymandias}\footnote{\href{https://github.com/Currodri/ozymandias}{https://github.com/Currodri/ozymandias}} a new cosmological simulation analysis package which allows for a smooth interfacing of halo and galaxy selection codes, in-depth analysis algorithms and simulated mock observations. It is developed with an easy-access philosophy in mind, combining {\sc Python} for accessing the data catalogues and Fortran 90, {\sc openMP} parallelised routines to access the raw \ramses~data. It takes a cosmological simulation snapshot and outputs a convenient galaxy and halo catalogue following the {\sc HDF5}\footnote{\href{https://www.hdfgroup.org/solutions/hdf5/}{https://www.hdfgroup.org/solutions/hdf5/}} format standard, which makes possible the study of many physical properties in the simulation without the need to read back again the original binary files. Additionally, these files are usually less than 0.3\% the size of the original outputs, making simple analysis and comparison between simulations much less memory and computationally expensive. {\sc Ozymandias} has been written to work with two different halo and galaxy finders (although the I/O for halo catalogues is fully compatible with other halo finders, providing the data structure information): the {\sc HaloMaker} code by \citet{Tweed2009}, a 3D density-threshold structure finder extended to use the shrinking spheres centring method \citep{Power2003TheStudy}; and the VELOCIraptor code \citep{Elahi2019a,Canas2019} in its enhanced version by \cite{Rhee2022PerformanceFormation}, a robust, scale-free, phase-space structure-finder. Once halos and galaxies are identified, a hierarchical structure is built between them, linking galaxies with their host halos and linking satellites with central galaxies. Then, in order to allow for an easy tracking of each galaxy/halo across cosmic time, particle lists for each identified object are compared between consecutive snapshots, allowing for an easy access to the progenitors in the previous snapshots. We use this package to identify the main galaxy, and refer to the \textit{galactic region} as the sphere with radius $0.2 R_{\rm vir,DM}$, $R_{\rm vir,DM}$ being the host DM halo virial radius. Quantities such as the galaxy stellar mass $M_*$ and gas mass $M_{\rm gas}$ are measured within this spherical volume at every snapshot.

\begin{table}
\centering
\caption{Runs of our new \nut~suite of simulations studying CR feedback included in this work. Columns indicate from left to right the simulation name, their employed solver and the details of the CR physics if CRs are present. All simulations use the same initial conditions, a MTT star formation prescription and the canonical configuration for mechanical feedback prescription. For simulations using the MHD or the CRMHD solver, all simulations employ a uniform comoving initial magnetic field of $B_0=3 \times 10^{-12}$~G.}
\label{table:nut_suite}
\begin{tabular}{m{0.15\columnwidth} m{0.1\columnwidth} m{0.6\columnwidth}} 
 \hline
 \textbf{Simulation} & \textbf{Solver} & \textbf{CR physics}\\ 
 \hline
  \textbf{HD} & Hydro & \xmark \\
  \textbf{MHD} & MHD & \xmark \\ 
  \textbf{CRMHD} & CRMHD & Streaming and $\kappa = 3\times 10^{28}$ cm$^2$ s$^{-1}$ \\ 
 \hline
\end{tabular}
\end{table}

\section{Results}\label{sec:results}

\subsection{The global effect of cosmic rays in the evolution of \nut}\label{subsec:global_effects}
\begin{figure*}
	\includegraphics[width=0.99\textwidth]{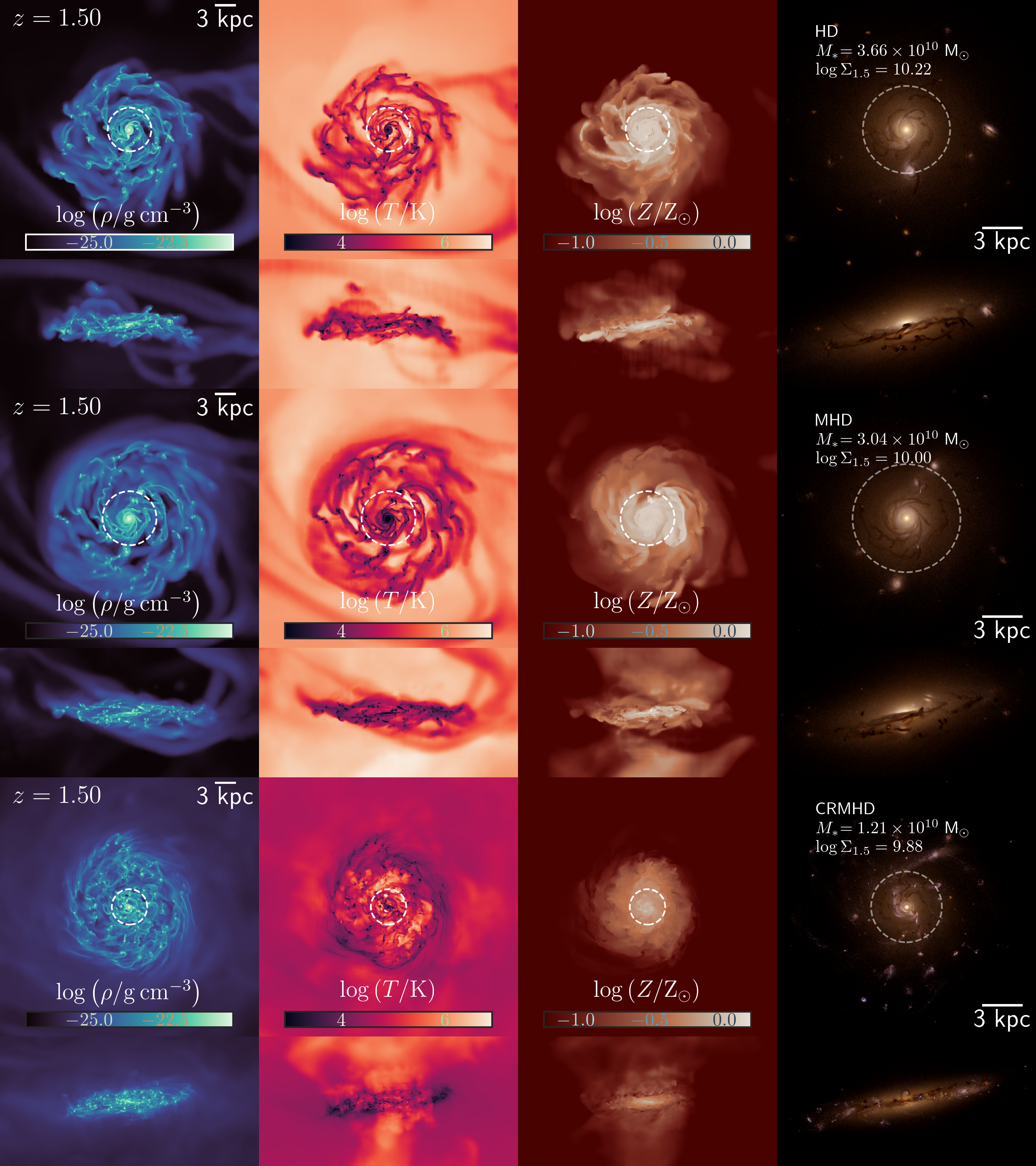}
    \caption{Projections of the \nut~galaxy for the three main models explored in this work: HD (top 2 rows), MHD (middle 2 rows), and CRMHD (bottom 2 rows), generated at $z=1.5$. For each model, the plot includes density-weighted projections of gas density (first column), gas temperature (second column), and gas metallicity (third column), 20~kpc on a side. Additionally, we present in the last column RGB synthetic {\sc Skirt} observations in the JWST F200W, F150W and F090W filters, respectively, for a zoom region of 10~kpc on a side. Face-on and edge-on views are shown for all models. Dashed white circles indicate the half-light radius $r_{50}$ obtained by fitting the RGB images with a 2D S\'ersic profile. For each simulation the stellar mass within \rgal\ and galaxy compactness $\log{\Sigma_{1.5}}$ \protect\citep{Barro2013CANDELS:2}, is listed, with a larger value indicating a more compact stellar emission. Compared to HD and MHD simulations, the disk of \nut~shows much more structure in the CRMHD simulation, dominated by a more diffuse ISM with less compact and dense clumps. The CGM is also strongly altered, with the cold dense filaments of HD and MHD replaced by a diffuse and cold halo more uniformly enriched by metals. The mock observations of HD and MHD show a galaxy dominated by an old stellar population and a large bulge, while the \nut~galaxy in CRMHD is transformed into a less compact system dominated by a disk. In this case, star forming regions are scattered across spiral arms and clumps all over the disk.}
    \label{fig:comp_init}
\end{figure*}

In Fig.~\ref{fig:comp_init} we show density-weighted projections of \nut~at $z=1.5$ for the three simulations (HD, MHD and CRMHD), comparing face-on views (line-of-sight parallel to the angular momentum vector of all baryons assigned to the galaxy by the halo finder) and edge-on views. We have also included mock JWST RGB images obtained with {\sc SKIRT} following the same parameter configuration as for Fig.~\ref{fig:canvas}. At $z = 1.5$, the \nut~galaxy has formed a thin gas disk in all simulations, that has large-scale spiral features with high density clumps and low density inter-arm regions. In the HD and MHD simulations (top and middle rows) the galaxy has the general properties of a disk galaxy within a virialised halo: a cold ($T<10^4$~K) thin disk with close to solar metallicity Z$_{\odot}$ is embedded within a low density, hot ($T>10^6$~K) CGM. The edge-on views aid in identifying long filaments of cold, dense and low metallicity gas inflowing into the galaxy. In the CRMHD simulation (bottom rows), while the dense and cold region of the disk can still be clearly seen in the projections, there are significant differences with respect to the no-CR simulations. Firstly, the gas disk shows much more filamentary substructure, and in comparison to the HD and MHD simulations the maximum value of density in the clumps is substantially reduced, as well as the volume occupied by low density, hot gas in the inter-arm regions. Secondly, the high density centre of the galaxy in HD and MHD extends to $\sim 1-2$ kpc, while in CRMHD it is $3$ times smaller. However, the most striking difference to the HD and MHD simulations is the thermodynamical state of the CGM: the CRMHD run has no clear inflow filaments, and it is instead dominated by homogeneous and denser gas at $T\sim 10^5$~K. Against this lower temperature `background', recent hot outflows can clearly be distinguished in the edge-on view, although the lower average metallicity of the disk gas compared to HD and MHD makes the metal enrichment of outflows less prominent. Instead, the CGM of \nut~is now more uniformly enriched to $\sim 0.1 $ Z$_{\odot}$. 

Furthermore, the mock JWST images of the HD and MHD simulations show \nut~ as a disk galaxy dominated by a massive bulge and an old stellar population, with dark, dense filaments seen as dust lanes in the edge-on views. Also, massive stellar clumps formed by large-scale disk instabilities persist due to inefficient feedback \citep[e.g.][]{Inoue2016Non-linearGalaxies,Bournaud2013THEREDSHIFT}. In comparison, the CRMHD simulation shows a dramatic bulge reduction, with the central luminosity of the galaxy reduced by 0.5 dex. Additionally, instead of having massive clumps within its disk, small star forming regions are seen all across the disk, with the luminosity of young stellar population more extended. To better quantify this important difference, we have measured the compactness of the stellar emission of \nut~using the total luminosity in the JWST filters. By fitting a 2D S\'ersic profile, we obtain half-light radii $r_{50}$ for the three face-on observations, which allow us to compute the compactness as defined in \citet{Barro2013CANDELS:2} and used for SDSS galaxies \citep{Baldry2020CompactRedshifts}:
\begin{align}
    \log \Sigma_{1.5} = \log(M_*/{\rm M}_{\odot}) - 1.5 \log (r_{50}/{\rm kpc})\,,
\end{align}
with a larger value of $\log \Sigma_{1.5}$ corresponding to a more compact galaxy. The HD simulation has the highest level of compactness, with the MHD simulation being slightly less compact but very similar as their total $M_*$ and $r_{50}$ are comparable. When considering the CRMHD simulation, a significantly reduced $r_{50}$ and a stellar mass three times smaller than in the HD simulation results in the least compact of our galaxies.

\begin{figure*}
	\includegraphics[width=\textwidth]{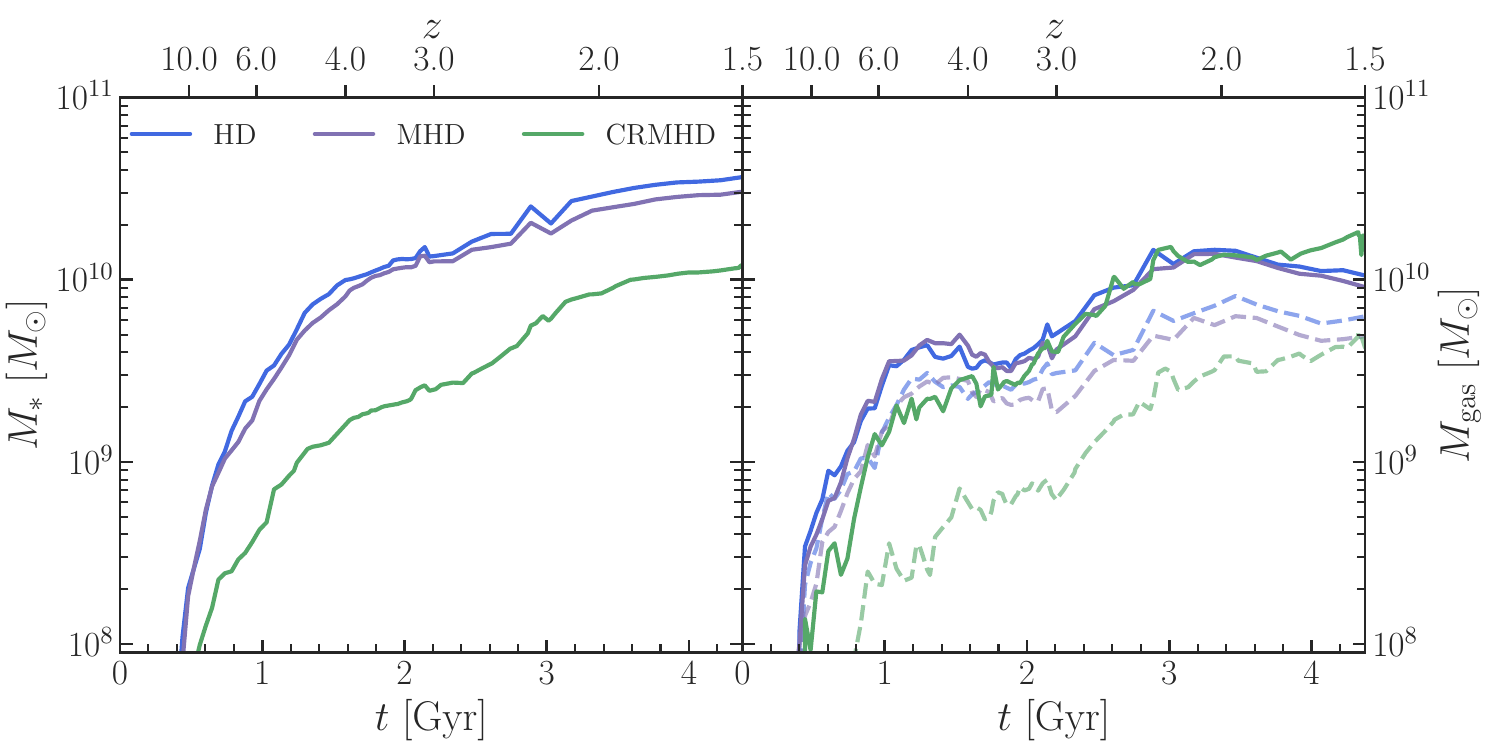}
    \caption{Evolution of the stellar (left panel) and gas (right panel) masses contained within the galactic region (i.e. $<0.2 R_{\rm vir, DM}$) of the \nut~galaxy in the HD (blue line), MHD (purple line) and CRMHD (green line) simulations. The right panel distinguishes the total gas mass (solid line) from the cold gas mass (dashed line), as defined by the constant entropy limit of the cold phase of the ISM by \protect\cite{Gent2012}. In all runs stellar mass grows rapidly up to $z\sim 4$, which marks the end of the accretion dominated phase, but the simulation with CRs has a stellar mass 1 dex lower than the HD and MHD runs. During this rapid growth period, the total gas mass of the CRMHD run is lower than in the other two simulations, but the largest difference is seen in the cold gas mass, which is reduced by $\sim 1$ dex at $z\gtrsim 4$. Below $z\sim 3$, the CRMHD run experiences a fast growth in cold gas mass, which translates into a faster growth of stellar mass at lower redshifts, causing the gas mass at $z=1.5$ to be similar to that in the HD and MHD runs, whilst the stellar mass remains 4 times lower.}
    \label{fig:stellar_mass}
\end{figure*}
\begin{figure}
	\includegraphics[width=\columnwidth]{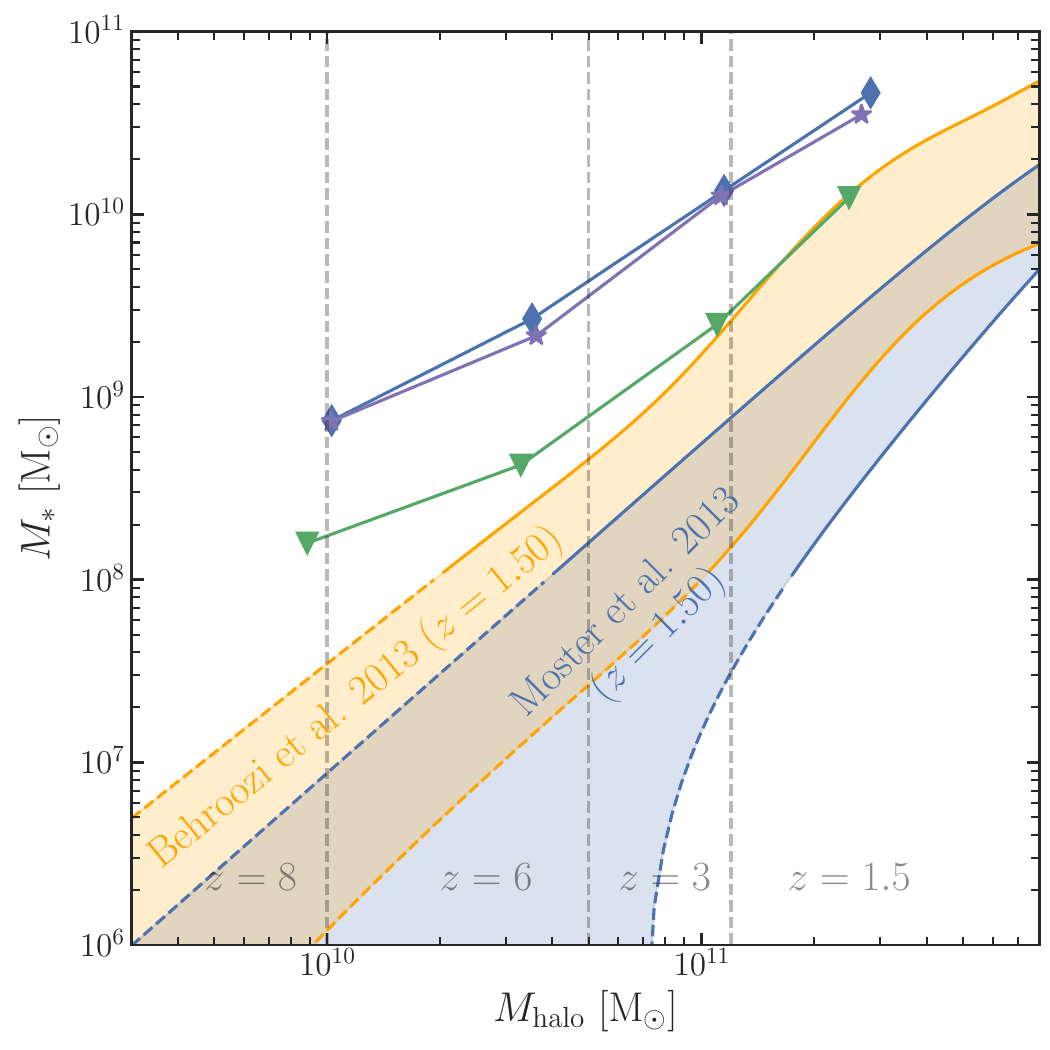}
    \caption{Stellar mass to halo mass function of the \nut~galaxy for our fiducial models: HD (blue), MHD (magenta) and CRMHD (green) shown in 4 redshift ranges ($z\sim 8,6,3$ and 1.5 -- separated by dashed vertical lines), representative of their evolution. We have superimposed, as a visual guide, the results from abundance matching of \protect\cite{Moster2013GalacticHaloes} (blue-shaded region) and \protect\cite{Behroozi2013The0-8} (orange-shaded region), evaluated at $z=1.5$. Coloured dashed orange and blue lines indicate the stellar masses at which observational mass functions are subject to incompleteness. The HD and MHD runs are able transform the majority of their baryonic content into stars, while the CRMHD run from early on ($z\sim 8$) sits $\sim 1$ dex below, approaching the prediction by \protect\cite{Behroozi2013The0-8} at $z\sim 1.5$.}
    \label{fig:smf}
\end{figure}

We begin by briefly revisiting the evolution of the HD and MHD runs, as these have already been explored for the same initial conditions in previous works \citep{Powell2011,Kimm2015,Martin-Alvarez2018}. The stellar (left panel) and gas (right panel) masses inside the galactic region are shown for the HD model with blue lines in Fig.~\ref{fig:stellar_mass}. The stellar mass growth follows the common trend for a MW-like halo \citep[e.g.][]{Kim2015,Martin-Alvarez2018}: 1) \textit{accretion phase}: significant gas accretion at high redshifts allows for a fast growth of stellar mass, reaching $10^{10}\msun$ at $\sim 1.5\,\rm Gyr$ after the start of the simulation; 2) \textit{feedback phase}: feedback from SNe and the formation of a virial shock after $z\sim 4$ slows down star formation, leading to self-regulation with $M_*\sim 4\times 10^{10} \msun$ at $z=1.5$, in agreement with the equivalent \textit{MFBm} model in \cite{Kimm2015}. The evolution of $M_{\rm gas}$ is also in agreement with the findings from \cite{Kimm2015}, exhibiting a rapid growth until $z\sim 5$. Subsequently, the fast depletion timescale reduces the gas mass inside the galactic region until $z \sim 3$. A new inflow of gas mass takes place from $z=3$ to $z=2$, coinciding with the last major merger that the \nut~galaxy experiences \citep[e.g.]{Martin-Alvarez2018}. Fig.~\ref{fig:stellar_mass} includes the MHD run with a primordial magnetic field of $3\times 10^{-12}$~G but no cosmic rays. This illustrates that the moderate magnetisation employed here does not significantly affect the global properties of the galaxy \citep{Su2017FeedbackFormation,Pillepich2018,Hopkins2020a,Martin-Alvarez2018}.

However, the addition of CRs gives rise to a radical difference in the star formation history (SFH) of our \nut galaxy: the steepness of the stellar mass growth is reduced during the accretion phase, resulting in $M_*\simeq 2\times 10^9\msun$, $\sim 1$ order of magnitude lower with respect to the HD and MHD models at $z\sim 4$. During this period of high-redshift gas accretion, the CRMHD run has only a $\sim \times 2-4$ deficit in $M_{\rm gas}$ inside the galactic region. Therefore the decrease in $M_*$ cannot be solely attributed to the depletion of $M_{\rm gas}$ from the galactic region. We also include on the right panel the cold gas mass (dashed lines) as defined by the constant entropy limit\footnote{The cold phase specific entropy limit is given by $s\leq \qty{4.4e8}{\erg\,\kelvin^{-1}\gram^{-1}}$, e.g. corresponding to a density of \qty{7.8e-23}{\gram\,\centi\metre^{-3}} at 300~K.} of \cite{Gent2012}, showing that for the CRMHD simulation the cold gas mass decreases by $\sim 1$ dex compared to HD and MHD at $z\gtrsim 4$. This points at the depletion of the cold gas mass in the ISM as the major cause of the lower stellar mass growth at these cosmic times. Following this period of strong star formation suppression, the CRMHD run exhibits a similar period of gas accretion beginning at $z\sim 3$ as the rest of the runs, which eventually translates into a new phase of steep $M_*$ growth. This growth of stellar mass is higher in relative terms for the CRMHD run than for the HD and MHD runs, reducing the gap between stellar masses to a factor of $\sim 4$, while the cold gas mass becomes similar in the three runs.

To provide further context for the evolution of the \nut~stellar masses, Fig.~\ref{fig:smf} shows the stellar mass ($M_*$) to halo mass ($M_{\rm halo}$) relation for the HD (blue diamonds), MHD (purple stars) and CRMHD (green triangles) runs at four different redshift ranges from $z=1.5$ to $z=8$. We define the halo mass as the full mass enclosed in gas and DM within $R_{\rm vir, DM}$ and the stellar mass of the galaxy. These redshift ranges are separated by vertical dashed lines with their central redshift values indicated at the bottom of the panel. Results from the abundance matching models by \citet{Behroozi2013The0-8} (orange shaded region) and \citet{Moster2013GalacticHaloes} (blue shaded region) at $z = 1.5$ are included for reference. The HD and MHD runs show very similar behaviour across the redshift range presented, with only a small stellar mass suppression by magnetic fields. Both feature large baryonic conversion efficiencies, close to the line of total conversion of gas into stars for a given halo, given by $f_{\rm b}M_{\rm halo}$, where $f_{\rm b}$ is the universal baryon fraction defined as $\Omega_{\rm b}/\Omega_{\rm m}$, the ratio of the cosmological baryon and matter density parameters. This result is consistent with the well established understanding of non-calibrated, canonical SN feedback: it fails to reproduce the results of abundance matching models, leading to a too massive stellar mass for a given dark matter halo mass. The HD run begins with $\sim 32$\% baryon conversion efficiency and grows to $\sim 56$\% by $z=2$, while the MHD run has a slightly lower efficiency of $\sim 47$\%. The CRMHD run shows a consistent decrease of the baryonic conversion efficiency across all redshifts, with a maximum value of $\sim 14$\% at $z=2$ and a minimum of 5\% at $z=6$. This significant decrease in stellar mass for a given DM halo in the CRMHD model results in a much better match to the semi-empirical constraints of \cite{Behroozi2013The0-8} by $z=1.5$. Despite the decrease of almost an order of magnitude in $M_*$ for the CRMHD run compared to the HD and MHD runs, the CRMHD model does not follow the `natural' extrapolation of the abundance matching models to lower halo masses. It is important to point out that reliable constraints on the number density of galaxies hosted by low mass halos of $10^{10} \msun$ between $z=6$ and $z=8$ are scarce, adding uncertainty in abundance matching models at high redshift. Additionally, the observational stellar mass functions used in these works are subject to observational incompleteness below $10^7 \msun$ (shown with the dashed orange and blue lines). We expect observations by JWST and Rubin-LSST will be able to remedy the situation in a near future.

\subsection{How cosmic rays alter the gas distribution and the star formation efficiency}\label{subsec:sf_efficiency}
\begin{figure*}
	\includegraphics[width=\textwidth]{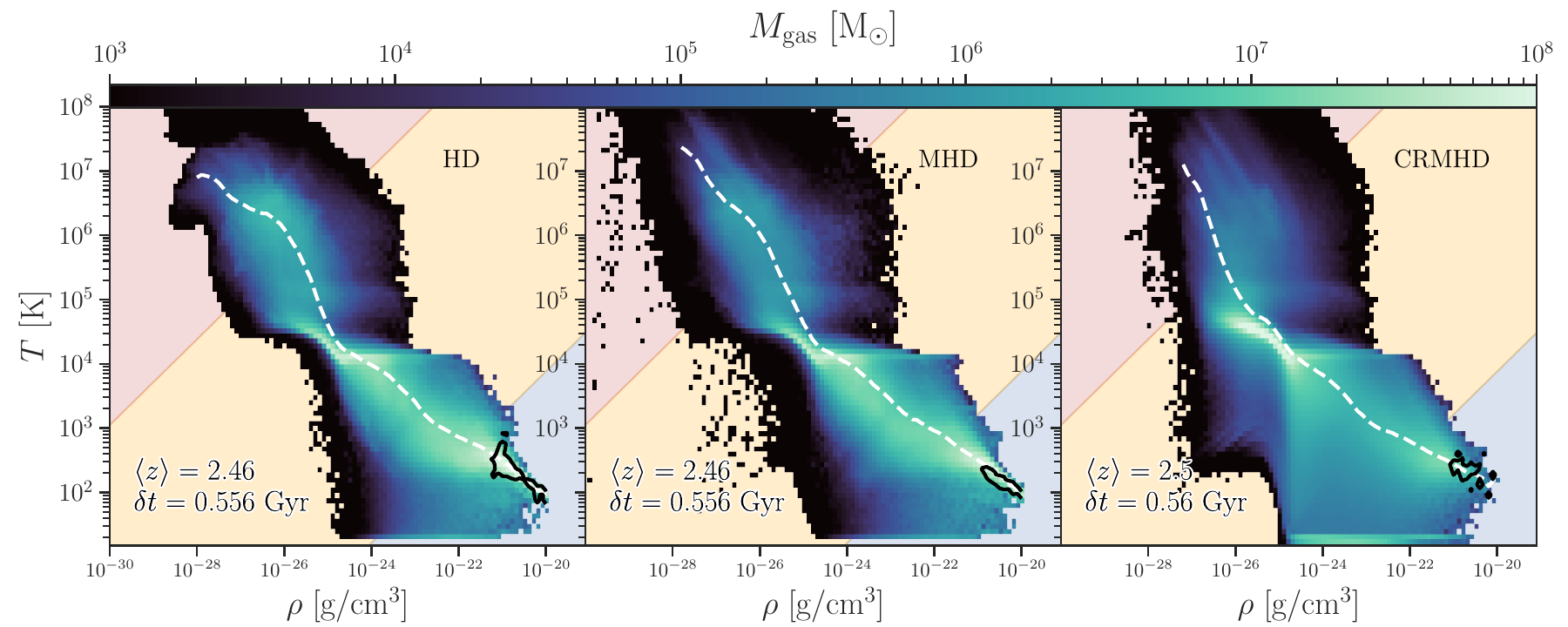}
    \caption{Stacked gas temperature-density phase diagrams of the \nut~galaxy ($<0.2R_{\rm vir,DM}$) in HD (left panel), MHD (middle panel) and CRMHD (right panel) at $z\sim 2.5$. Colour maps indicate the total gas mass in each bin, with lower masses shown in dark blue and higher masses in bright aquamarine. We include a dashed white line corresponding to the median of the 2D profile. Gas phases are separated by lines of constant specific entropy given by \protect\cite{Gent2012}, into hot (top left, red), warm (central, orange), and cold (bottom right, blue). Black contours indicate the regions where $\sim 80$\% of the gas with $\epsilon_{\rm SF}^{\rm MTT} \geq 1$\% lie. These show how the CRMHD run has much lower total gas mass in regime favourable to star formation.}
    \label{fig:pd_mass_galaxy}
\end{figure*}
As shown in Section~\ref{subsec:global_effects}, the CRMHD simulation contains a factor of $\sim 2-3$ lower amount of gas in the galactic region of \nut~(compared to HD and MHD) between $z=8$ and $z=4$. This epoch also is where we find the largest discrepancy in $M_*$ between the CRMHD and HD/MHD runs. However, the 10 times smaller stellar mass measured at these redshifts cannot be solely attributed to the lower gas content in the galaxy which is much more moderate, and we need to consider whether the star formation efficiency of the gas is reduced compared with the HD and MHD runs. To explore the physical conditions of the ISM gas in \nut, Fig.~\ref{fig:pd_mass_galaxy} shows temperature-density phase diagrams for the three runs used in this study. Phase diagrams based on a single simulation snapshot can be hard to interpret as they are subject to the stochasticity driven by processes such as feedback events and mergers. To avoid this issue, the phase diagrams in Fig.~\ref{fig:pd_mass_galaxy} are averaged over multiple snapshots within the galaxy's dynamical time at $z\sim 2.5$, weighted by their difference in time to $z \sim 2.5$. We estimate the galaxy's dynamical time $\tau_{\rm dyn}$ as the time required for a test particle to complete one full orbit at $r = 0.2 R_{\rm vir,DM}$ with circular velocity:
\begin{align}
    v_{\rm circ}(r) = \sqrt{\frac{G M(<r)}{r}}\,,
\end{align}
where $M(<r) = M_{\rm b} + M_{\rm DM}$ is the combined baryon and DM mass within that sphere. Additionally, to ensure the comparison between runs occurs over the same time span, we use the largest $\tau_{\rm dyn}$ of the three runs compared, which for $z\sim 2.5$ corresponds to the CRMHD simulation and has a value of $\delta t = 0.556$~Gyr. Therefore, our measurements have a central redshift $\langle z \rangle$ and a physical time range $\delta t$. 

For the HD run, the left panel of Fig.~\ref{fig:pd_mass_galaxy} shows that the low density gas (\qtyrange[print-unity-mantissa = false]{e-26}{e-24}{\gram\,\centi\metre^{-3}}) in the galactic region is dominated by gas with temperature $T \sim 10^4$~K (where hydrogen recombination and resonance lines of atomic hydrogen dominate the cooling of gas) and above, a typical signature of SNe shock-heating. Gas accumulates in the $T \sim 10^4$~K region due to thermal instability \citep{Field1969Cosmic-RayGas,Wolfire1995TheMedium}. When this gas is perturbed by processes such as shock compression, it collapses into cold dense clumps that populate the tip of the phase diagram in the bottom right corner. In agreement with the results found by \cite{Martin-Alvarez2018,Martin-Alvarez2020} for \nut, the addition of magnetic fields helps sustain a somewhat larger fraction of gas in hotter phases, as well as allows for gas cells to reach lower densities whilst retaining their temperature. This is reflected in the size of the `dark' regions of the phase diagram with densities below $10^{-24} \,\rm g\,cm^{-3}$, which are larger in the MHD run as opposed to the HD run, due to the extra magnetic pressure support. To further aid this analysis, we have separated the temperature-density diagrams into three phases according to specific entropy $s$: cold and dense (blue region), warm (orange region) and hot and diffuse (red region). This separation is inspired by the work of \cite{Gent2012}. Similar amounts of gas are found in the hot and warm regions of the HD and MHD runs, with the MHD run having a slightly higher gas mass in the warm phase (from 49\% in HD to 55\% in MHD). The hot phase only contributes 3\% to the total gas mass in both HD and MHD. 

However, the gas phase diagram for the CRMHD run (right panel of Fig.~\ref{fig:pd_mass_galaxy}) shows marked differences compared with the non-CR models: (a) at densities $\rho < 10^{-24}\,\rm g\, cm^{-3}$, gas is no longer confined to temperatures above $10^4$~K when $\rho < 10^{-24}\,\rm g\, cm^{-3}$, which opens a new region in phase space for low density cold gas; (b) the bottom right of the warm region (i.e. $T<10^4$~K with $10^{-24}< \rho < 10^{-22}\,\rm g\, cm^{-3}$) has a larger scatter around the median (dashed white line) of the 2D histogram; and (c) the cool dense region below the diagonal blue line which separates the cold and warm phases contains less gas mass than the HD and MHD runs. In fact, when computing the total masses in the three phases, the cold phase mass in the CRMHD run is reduced by a factor of two compared to the HD model as this gas now resides in the warm medium. Notably, it is in these dense regions of the phase diagram that we expect gas to undergo a more efficient star formation. This considerable reduction of cold gas mass explains, at least partially, the causal link with the measured decrease of stellar mass in this simulation as inferred from Fig. \ref{fig:stellar_mass}. Note that we have not yet quantified the turbulent or magnetic support of the gas, and how these affect the efficiency of star formation. Indeed, in combination with the thermal pressure, the additional CR pressure is able to support gas against further collapse, and the build up of a pressure gradient allows for the warm gas reservoir above the disk to be kept there for longer \citep[e.g.][]{Dashyan2020,Chan2022TheGalaxies,Farcy2022Radiation-MagnetoHydrodynamicsGalaxies,Armillotta2022Cosmic-RayEnvironments}. This depletion of cold gas, now transferred to the warm phase in the presence of CRs is in good agreement with other works \citep[e.g.][]{Dubois2019,Nunez-Castineyra2022Cosmic-rayLuminosities}. By considering the medians of the 2D phase diagrams, two main features are readily visible: (i) the larger contribution to the mass budget by gas between \qtyrange[print-unity-mantissa = false]{e-24}{e-22}{\gram\,\centi\metre^{-3}} in the CRMHD run creates a more pronounced concave inflection point towards lower $T$ at $\rho \sim \qty[print-unity-mantissa = false]{e-23}{\gram\,\centi\metre^{-3}}$; (ii) gas mass accumulates more tightly around the thermal instability shoulder at densities below \qty{e-24}{\gram\,\centi\metre^{-3}}, driving the median line closer to this feature of the phase diagram as the relative amount of gas above $T \geq \qty[print-unity-mantissa = false]{e5}{\kelvin}$ is greatly reduced, particularly at higher densities.

\subsection{Cosmic rays reshape how supernovae interact with the galactic region}\label{subsec:sn_outflows}
\subsubsection{Distribution of supernova sites and feedback effectiveness}\label{subsubsec:sn_sites}
\begin{figure}
	\includegraphics[width=\columnwidth]{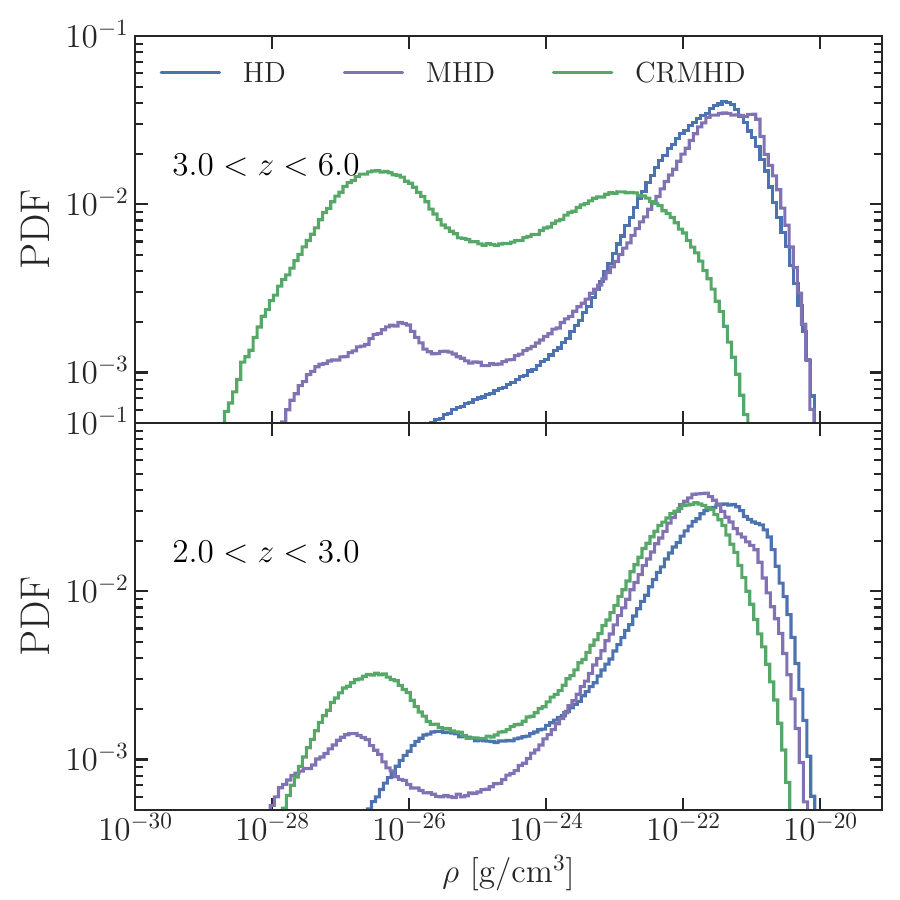}
    \caption{Distribution of gas densities in which SN events take place in the redshift range $z=6.0-3.0$ (top panel) and $z=3.0-2.0$ (bottom panel). Data corresponding to the CRMHD simulation is shown in green, the MHD simulation in purple and the HD simulation in blue. Across all redshifts, the majority of SN events in the HD and MHD simulations are found at high densities (\qtyrange[print-unity-mantissa = false]{e-22}{e-21}{\gram\,\centi\metre^{-3}}), in which SNe explosions suffer from strong radiative cooling. Contrarily to HD and MHD, in the CRMHD simulation at early times (top panel) the distribution is strongly bimodal, with a peak at low densities $\sim \qty[print-unity-mantissa = false]{e-27}{\gram\,\centi\metre^{-3}}$ and a peak at $\sim \qty[print-unity-mantissa = false]{e-23}{\gram\,\centi\metre^{-3}}$. SN events taking place at lower densities during this period are more effective in suppressing star formation. This bimodality for CRMHD is still present in the lower redshift range (bottom panel), but the contribution of the low density peak is significantly suppressed, with the overall distribution resembling more the one obtained for the HD and MHD simulations.}
    \label{fig:sne_pdfs}
\end{figure}
The modifications we observed to the gas distribution across ISM phases also have an effect on the distribution of SN explosions across the multi-phase ISM. These local properties will affect the efficiency of SN feedback on self-regulating star formation and modifying the ISM properties. We explore this issue by recording the local ISM conditions for each SN event. In Fig.~\ref{fig:sne_pdfs} we plot the distribution of SN site gas density for two redshift ranges: between $3<z<6$ (top panel), i.e. the period of largest $M_*$ difference between CRMHD and HD/MHD, and between $2<z<3$ (bottom panel), which samples the period of fast growth of $M_*$ in CRMHD. We find for the HD simulation that (as in previous works \citealp{Kimm2015}) these PDFs are intrinsically bimodal with the two peaks at low and high densities, characterising the hot and cold phases of the ISM \citep[see e.g.][]{Gent2012}. However, the high density peak dominates the overall distribution, which at higher redshifts (top panel) resembles more a log-normal distribution of the cold ISM. When SNe explode in the HD case, young star particles have not had the chance to leave the dense regions of the galaxy in which they were born. This means the $10^{51}$ erg from a single SN\footnote{At high densities the Sedov-Taylor phase will hardly be captured by our $\sim 20$ pc resolution, so the mechanical feedback method from \cite{Kimm2017Feedback-regulatedReionisation} will mainly inject the momentum reached at the end of the Sedov-Taylor phase.} will be deposited into a larger gas mass per cell, hence reducing the maximum launch velocity and making it harder for the outflowing gas to escape the gravitational potential. 

Including magnetic fields in the MHD simulation moves the low density peak of the distribution to lower densities ($\sim \qty[print-unity-mantissa = false]{e-27}{\gram\,\centi\metre^{-3}}$) and produces a more bimodal distribution at higher redshifts. As already implied by the phase diagrams of MHD (see Fig.~\ref{fig:pd_mass_galaxy}), the additional magnetic support alters the gas phase distribution inside the galaxy, allowing for SNe to take place in lower density gas. Furthermore, we also attribute this peak at lower densities to the increased clustering of star-forming regions \citep[e.g.][]{Hennebelle2014,Hennebelle2022InfluenceClump}, as magnetic fields reduce the cold gas fragmentation, making feedback events more correlated to each other and hence promoting the formation of super-bubbles. 

The presence of this bimodality in SN sites is strongly altered for the CRMHD simulation compared to no-CR runs. Although the effects seen in the MHD simulation, which arise from the support of magnetic fields, should also be in place in the CRMHD simulation, we have already seen (in Fig.~\ref{fig:pd_mass_galaxy}) that the presence of CR pressure and heating mechanisms dominates over the effect of magnetic fields. The ISM in the CRMHD run is strongly depleted of high density cold gas, with the high density peak of the bimodal distribution now at $\sim \qty[print-unity-mantissa = false]{e-23}{\gram\,\centi\metre^{-3}}$ in the CRMHD run instead of $\sim \qty[print-unity-mantissa = false]{e-21}{\gram\,\centi\metre^{-3}}$ in the HD and MHD runs. The regions of star formation (black contours) in Fig.~\ref{fig:pd_mass_galaxy} are pushed to lower densities, and the warm phase dominates the gas mass budget in the galactic region. Therefore, it is more common for SN explosions to take place at lower densities ($\leq 10^{-22}\,\rm g\,cm^{-3}$), with a large number of SNe exploding at $\rho \sim 10^{-27}\,\rm g\, cm^{-3}$. We attribute this to star formation happening in a less clustered form throughout the galactic region, with the consequent feedback events being able to sample lower density regions shaped by CR pressure and previous SN-driven bubbles. In addition, SN remnants evolve for longer and maintain higher temperatures past the canonical evolutionary sequence \citep[e.g.][]{Naab2016} when the effects of CR pressure are taken into account (see \citealt{RodriguezMontero2022MomentumRays} for a parameter space study of CR feedback effects on the evolution of individual SN using 3D MHD simulations). Furthermore, similarly to the behaviour of fully thermal SN remnants \citep{Thornton1998}, when CRs are accounted for in the evolution of SN explosions in low density environments, they can reach a larger value of deposited momentum compared to high density environments \citep{RodriguezMontero2022MomentumRays}.

The bottom panel of Fig.~\ref{fig:sne_pdfs} shows the PDF of SN site densities instead between the redshifts of 3 and 2. In this lower redshift range we find that the CRMHD PDF begins to converge to the ones obtained for the HD and MHD runs, although maintaining a higher fraction of SNe happening at lower density (i.e. $\sim \qty[print-unity-mantissa = false]{e-27}{\gram\,\centi\metre^{-3}}$). \nut~transitions into a rotationally supported (i.e. ratio of rotational velocity to dispersion velocity larger than unity) disk at $z\sim 3.5$ \citep{Martin-Alvarez2020} in HD and MHD simulations, while we found that for our CRMHD run this transition occurs closer to $z\sim 2.3$. The \nut~galaxy is dominated by the gas velocity dispersion between $z=6-3$, which means that CRs released by SN events within the galactic region can easily diffuse across the ISM and affect additional star-forming regions. However, in the lower redshift range of the bottom panel we are instead capturing the morphological transformation into a dynamically cold disk. With a disk morphology in place, CRs will be more prone to escape perpendicular to the disk, having a less drastic effect in the cold phase of the ISM at lower redshifts. We predict a dual behaviour depending on the morphology of the galaxy, which we will explore in future work.

\subsubsection{Influence of cosmic rays in the launching of outflows}\label{subsubsec:outflows}
The CRMHD simulation has both lower $M_{\rm gas}$ and $M_*$ at high redshift than the HD and MHD models. This must be, at least partially, due to differences in the outflow and inflow rates of these simulations. In this work we focus on the analysis of outflowing gas. We measure instantaneous outflow rates $\dot{M}_{\rm outflow}$ as
\begin{align}
    \dot{M}_{\rm outflow} (r) = \frac{1}{\delta r_{\rm shell}}\sum_{i}\rho_i v_{r,i} (\Delta x_i)^3\,,
\end{align}
where we sum over all cells with positive radial velocities $v_{r,i}$ in a thin spherical shell of width $\delta r_{\rm shell}$ at a distance $r$ from the galactic centre. The width of the shell is taken to be $0.01\, R_{\rm vir,DM}$, which provides a consistent measure across simulations and cosmic time. Each cell is characterised by its gas density $\rho_i$, the radial velocity $v_{r,i}$ and the cell width $\Delta x_i$. Previous studies with the \nut~galaxy \citep{Tillson2015} found the contribution to the inflowing gas from satellites to be small ($\sim 1\%$) compared to that from filaments. Nevertheless, for the analysis of the outflowing gas, satellites that cross our thin shell may contain gas that could appear to be flowing radially outward in the frame of the central galaxy (either due to satellite rotation, or because a given satellite crosses the measuring shell on its way back to the outer halo regions after its apocentric passage). To avoid these subtleties, we subtract all contributions in the thin shell from passing substructures. We thus remove gas located within the tidal radius of DM sub-halos containing more than 1000 particles. The tidal radius is estimated as the Jacobi radius with an isothermal sphere approximation for DM halos given by Eq. 8.107 of \cite{Binney2008GalacticDynamics}. Finally, we divide outflowing gas into separate temperature phases as follows:
\begin{itemize}
    \item Hot gas: $T>10^5$~K,
    \item Warm ionised gas: $10^5>T>9000$~K,
    \item Warm neutral gas: $9000>T>1000$~K and
    \item Cold gas: $T<1000$~K\,.
\end{itemize}
\begin{figure}
	\includegraphics[width=\columnwidth]{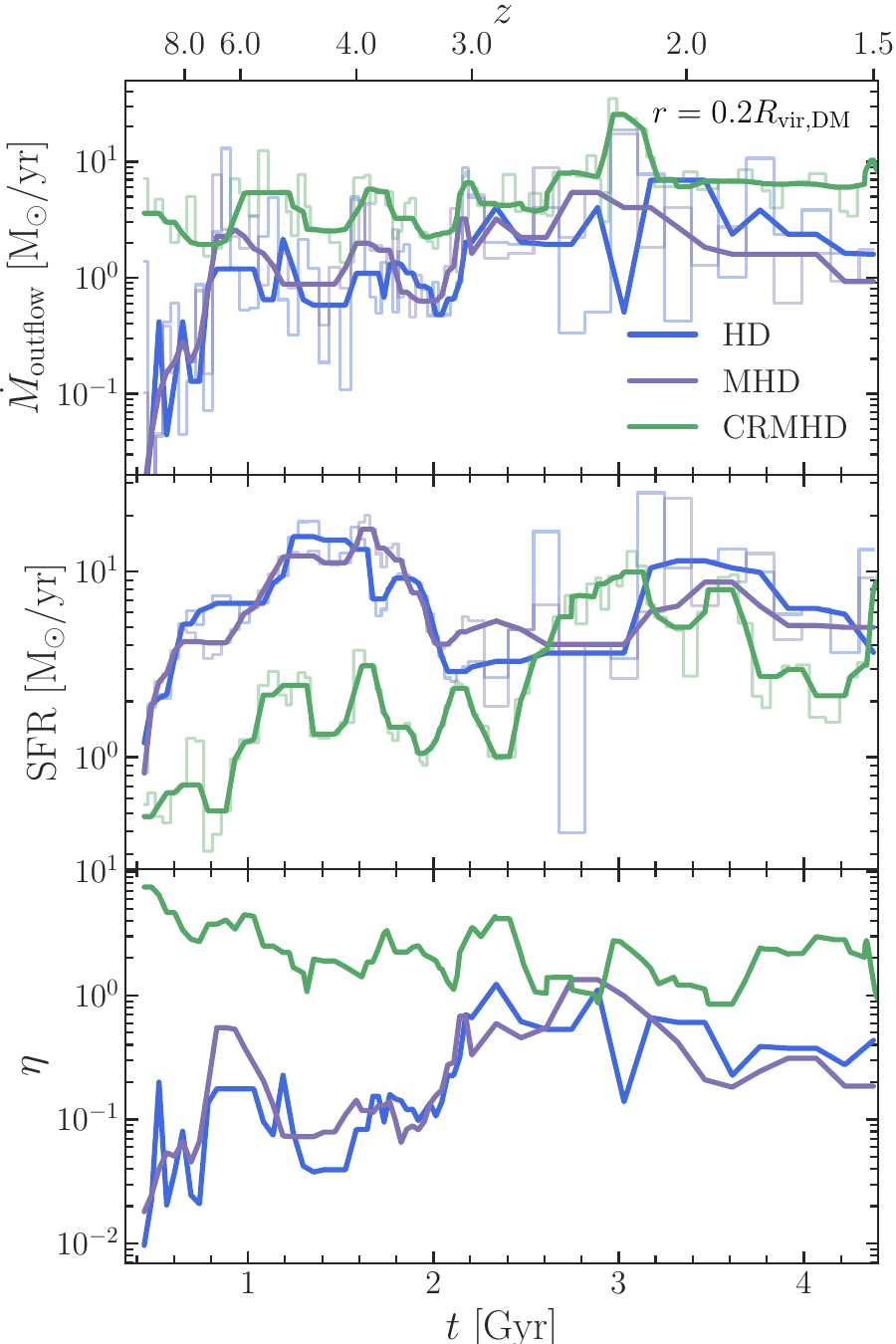}
    \caption{\textbf{Top panel:} Outflow rate measured in a thin spherical shell at \rgal\ for the HD, MHD and CRMHD runs. The histograms show the instantaneous mass-flow rate through the thin shell, while the solid lines are the running median of these measurements using a filter of size 200 Myr. \textbf{Middle panel:} Star formation rate (SFR) averaged over time bins of 100 Myr. \textbf{Bottom panel:} Mass-loading factor $\eta$ computed using the running median results of $\dot{M}$ and SFR shown in the top panels. The simulation with CRs has large outflow rates, $\times 10-100$ higher than the HD and MHD simulations, and maintains them across all cosmic time explored. Given the large suppression of star formation for $z\geq 3$, the CRMHD simulation has outflow mass-loading consistently exceeding unity, only reached by the HD and MHD simulations during the merger-induced starburst at $z\sim 3-2$.}
    \label{fig:instantaneous_outflow_20rvir}
\end{figure}

In Fig.~\ref{fig:instantaneous_outflow_20rvir} we show the results of this outflow analysis in a thin spherical shell at \rgal\ for the three simulations across $\sim 4.5\,\rm Gyr$ of evolution, where $\dot{M}_{\rm outflow}$ (top panel), star formation rate (SFR) averaged over $100 \, \rm Myr$ (middle panel), and the mass loading factor $\eta=\dot{M}_{\rm outflow}/{\rm SFR}$ (bottom panel) is shown. Solid lines are the result of a running median smoothing with a size of 200 Myr. Note that we do not restrict our analysis to particular assumptions about the possible time delay between a star formation event in the galaxy and the consequent outflow at \rgal. We distinguish three distinct time phases when comparing the HD, MHD and CRMHD simulations: i) between $z\sim 13$ and $z\sim 7$ CRMHD reaches an outflow rate of $2-4 \,\rm \msun\,yr^{-1}$, whereas the HD and MHD runs are closer to $0.1-0.4\,\rm \msun\,yr^{-1}$; ii) between $z\sim 7$ and $z\sim 3$ the outflow rates of the HD and MHD runs reach $1 \,\rm \msun\,yr^{-1}$ (with the MHD run having overall larger values than HD), during times of strong starbursts. However, given their average SFR of $10 \,\rm \msun\,yr^{-1}$, their mass loading factors remain well below unity, while the CRMHD simulation consistently exceeds $\eta \sim 1$ during starbursts; and iii) between $z\sim 3$ and $z\sim 1.5$ the SFR of the three runs peaks at $z\sim 2.5$ at a value of $10 \,\rm \msun\,yr^{-1}$ (and slowly drops to $2-4 \,\rm \msun\,yr^{-1}$). After this time, only CRMHD maintains a steady outflow rate of $8-9 \,\rm \msun\,yr^{-1}$ while MHD drops to $2 \,\rm \msun\,yr^{-1}$ and HD is kept at $2-4 \,\rm \msun\,yr^{-1}$, with a jump to $8 \,\rm \msun\,yr^{-1}$ around $z\sim 2$. It is also important to note that in comparison with the HD and MHD runs, the CRMHD runs experiences a fast rise in SFR of 1 order of magnitude between the redshifts of 3 and 2.3. As mentioned before, during this period \nut~experiences the merger with a large companion, causing a large increase in the cold gas mass of the galaxy (see Fig.~\ref{fig:stellar_mass}) and a burst in star formation. We find that this period is connected with a large increase of inflow rate to \nut~in the CRMHD run (Rodr\'iguez Montero et al., in prep.), caused by the cold, high density extraplanar gas supported by CR pressure to become unstable during the merger interaction. Overall, the cosmic evolution of $\dot{M}_{\rm outflow}$ appears more steady than the SFR for the CRMHD run. This suggests that while SN feedback is rarely efficient in the HD and MHD runs with large amount of gas consumed into stars, in the CRMHD run large quantities of gas are ejected from the galaxy and gas-to-stars consumption rate is much lower.

\begin{figure}
	\includegraphics[width=\columnwidth]{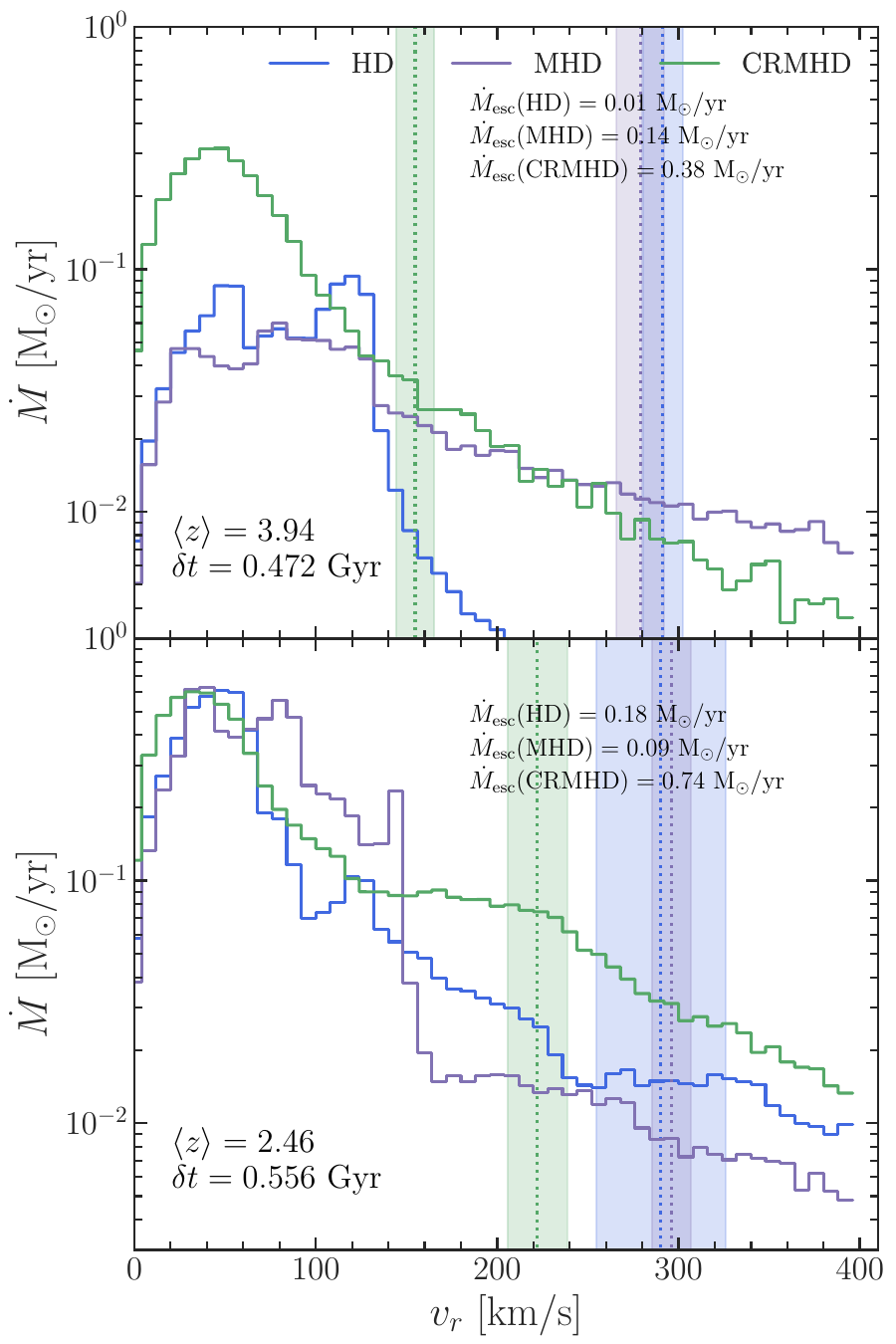}
    \caption{Histogram of mass outflow rates as a function of gas radial velocity measured in a thin shell at $0.2 \,R_{\rm vir, DM}$ and $\langle z\rangle = 3.94$ (top panel) and $\langle z\rangle = 2.46$ (bottom panel). We compare the HD (blue), MHD (purple), and CRMHD (green) simulations, stacking the histogram values in snapshots spanning a $t_{\rm dyn}$ range, using the method detailed in Section~\ref{subsec:sf_efficiency}. Vertical dotted lines show the stack-weighted averages $v_{\rm esc}$ for the redshift quoted, with the corresponding standard deviations denoted by the shaded regions. Values for the total mass outflow rate above $v_{\rm esc}$ ($\dot{M}_{\rm esc}$) are included in each panel for each simulation. The peak of the outflowing gas distribution is both more prominent and displaced to lower radial velocities in the CRMHD simulation in both redshift ranges compared to HD and MHD. At high redshift, the shallower gravitational potential in the CRMHD simulation allows a much larger fraction of the high velocity tail to exceed the local $v_{\rm esc}$.}
    \label{fig:PDF_flowrate_comp}
\end{figure}
In order to further understand the underlying differences between outflows in the non-CR runs and CRMHD, we compare their outflow velocity distributions during two distinct phases of the SFH: 1) a peak in SFR present in all simulations at $z\sim 4$; and 2) a period of similar SFRs across all runs ($\sim 8-10 \,\rm \msun\,yr^{-1}$) occurring approximately at $z\sim 2.5$. The latter is characterised by a merger with a small companion. We compute stacked mass-flow rate histograms following the same method detailed in Section~\ref{subsec:sf_efficiency} for weighted stacking of PDFs, employing snapshots within a dynamical time of the galaxy. These histograms show the total $\dot{M}$ separated into radial velocity bins for the same thin shell employed in Fig.~\ref{fig:instantaneous_outflow_20rvir} at \rgal. We show the HD, MHD and CRMHD simulations in blue, purple and green, respectively. The top panel of Fig.~\ref{fig:PDF_flowrate_comp} shows that for $z=\langle 3.94\rangle$ the outflowing gas carrying the majority of the mass outside of the galaxy is found at lower radial velocities in the CRMHD run than in HD and MHD. While the HD distribution is dominated by gas between 60$\, \rm km\,s^{-1}$ and 120$\, \rm km\,s^{-1}$, it rapidly drops towards lower velocities, with a small fraction of the total outflow rate ($0.05 \,\rm \msun\,yr^{-1}$) exceeding the $v_{\rm esc}$ of the galaxy (given by the vertical dotted blue line). This escape velocity is computed for each time $t$ by counting the total mass enclosed within \rgal. The MHD simulation instead shows a tail that extends to higher velocities, contributing 10 times more to the mass outflow rate above $v_{\rm esc}$ than HD. However, despite this higher efficiency driving fast outflows, the majority of the outflowing mass between 60$\, \rm km\,s^{-1}$ and 120$\, \rm km\,s^{-1}$ remains the same, and the mass above $v_{\rm esc}$ is 3 times lower than the CRMHD. This is due to the fact that $v_{\rm esc}$ is half the value for CRMHD than for HD/MHD, a signature of a shallower potential well. This is due to the CR feedback significantly affecting the growth of the central stellar mass of \nut~at high redshift. Therefore, the larger outflow rates of CRMHD at $z\sim 4$ compared to HD and MHD can be explained by a larger fraction (almost 1 dex compared to HD and MHD) of `low-velocity' gas and an easier escape of fast outflows in a less deep gravitational potential. To further explore these trends, we also analyse the outflow velocity distribution at $z\sim 2.5$ (see bottom panel of Fig.~\ref{fig:PDF_flowrate_comp}). Although now the fast growth of \nut~in the CRMHD simulation between $z=3$ and $z=2$ somewhat bridges the gap in $v_{\rm esc}$, this simulation still has 6 times as much gas outflowing above $v_{\rm esc}$ than the HD and MHD runs. This is not surprising given that during this period $\dot{M}_{\rm outflow}$ in CRMHD is more than an order of magnitude higher than HD and MHD. Furthermore, these plots also suggest that the majority of the outflowing gas at \rgal\ lacks the kinetic energy necessary to escape the local gravitational potential in all simulations. However, it remains an open question whether this gas needs to have the energy necessary to escape the halo in order to reduce the reservoir of gas available for star formation or if it is enough to prevent its re-accretion over sufficiently long timescales \citep[see e.g.][for preventive feedback in the context of AGNs]{Somerville2015}. As far as our current analysis goes, the CRMHD model is able to reach higher mass loadings than the no-CR runs. These outflows are mainly comprised of a slow moving wind that, without any further acceleration beyond $0.2 R_{\rm vir, DM}$, will not be able to escape the halo and will eventually fall back into the galaxy if undisturbed by future feedback events.

\subsubsection{Dynamics of multi-phase outflows}\label{subsubsec:phases}
\begin{figure*}
	\includegraphics[width=\textwidth]{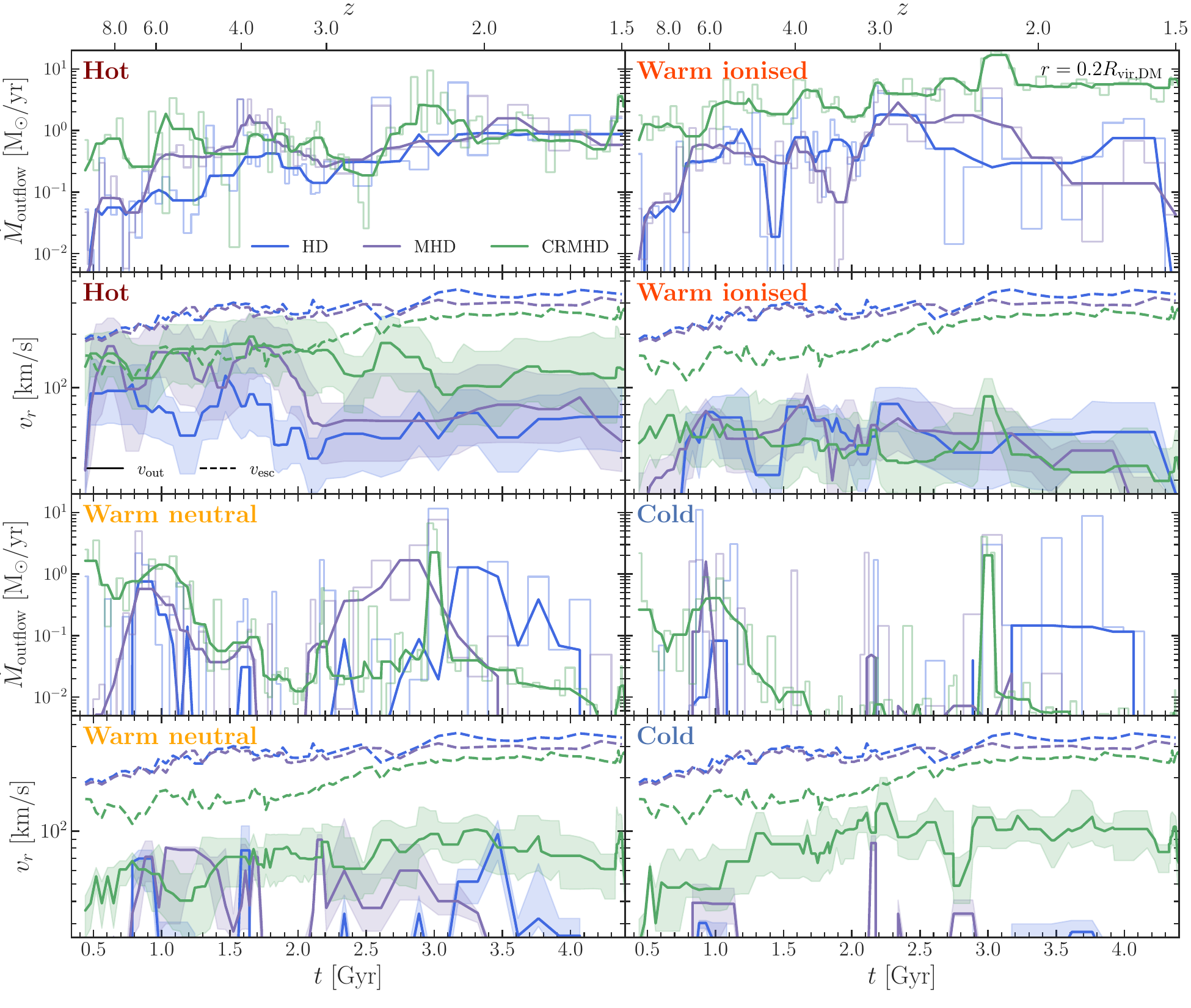}
    \caption{Outflow budget and kinematics of hot (top left pair), warm ionised (top right pair), warm neutral (bottom left pair) and cold (bottom right pair) phases for the HD (blue), MHD (purple) and CRMHD (green) runs measured at \rgal. Each pair of panels show the instantaneous outflow rate (top) and the $\dot{M}_{\rm outflow}$-weighted radial velocity of outflowing gas (bottom). The $\dot{M}_{\rm outflow}$ panels show the same quantities as Fig.~\ref{fig:instantaneous_outflow_20rvir}: low opacity histogram is the instantaneous value, while the solid lines show the running median with a filter of $200$~Myr. In the $v_r$ panels, the escape velocity of each model at the shell radius is showed with dashed lines, for ease of comparison to the median $v_{\rm outflow}$. Shaded regions represent the second and fourth quartiles of the underlying weighted distribution. Hot gas attains the largest radial velocities of all phases, with an almost constant value of $\sim 100\, \rm km\,s^{-1}$ for $z\geq 3$ and $\sim 70\, \rm km\,s^{-1}$ for $z\lesssim 3$, dominating the outflow rate in HD and MHD simulations. The hot phase in CRMHD reaches the largest outflow velocities ($\sim 150-200\, \rm km\,s^{-1}$ for $z\geq 2.5$ and $\sim 100\, \rm km\,s^{-1}$ for $z\lesssim 2.5$), with a non-negligible fraction of the outflowing gas mass exceeding $v_{\rm esc}$ consistently up to $z\geq 2.5$. The dominant mass-flow rate is coming from the warm ionised phase in the CRMHD, while above $z \sim 5$, warm neutral and cold phases largely contribute to the total outflow rate. Warm neutral and cold phases are scarce in the HD and MHD simulations, usually identified as low velocity peaks which can matched with crossing gas features of orbiting satellites. However, in the CRMHD run these outflows while slower than the hot gas can still reach $\sim 100\, \rm km\,s^{-1}$ at lower redshifts.}
    \label{fig:outflow_phases_20rvir}
\end{figure*}
The higher outflow rates measured for the CRMHD simulation despite a smaller SFR are a consequence of the energy deposited by SNe being more efficient at expelling gas from the galactic disk. To provide evidence for this higher efficiency causing these higher mass-loading results in CRMHD, Fig.~\ref{fig:outflow_phases_20rvir} shows the $\dot{M}_{\rm outflow}$ and $\dot{M}_{\rm outflow}$-weighted radial velocities of outflowing gas $v_{\rm outflow}$ in the HD (blue), MHD (purple) and CRMHD (green) simulations, separated in the different phases we explore. HD and MHD outflows are dominated by the hot and ionised phases during the majority of the evolution of \nut. Outflowing gas in the warm neutral and cold phases in these runs are identified with sudden, low velocity peaks in the velocity plots, which we found to be related to clumps of gas in eccentric orbits around \nut, rather than genuine outflows. Instead, the CRMHD run shows that the hot phase never dominates the outflowing mass rate. The warm ionised phase dominates ($1-10\,\rm \msun\,yr^{-1}$) the outflowing gas mass up until $z\sim 4$, with considerable contributions of warm neutral and cold gas at higher redshifts. We find that multi-phase outflows ($\sim$1:2:4:2 ratios for cold, warm neutral, warm ionised and hot phases, respectively) are only found consistently in the CRMHD simulation, and at high redshift ($z\geq 4$). The HD and MHD runs fail to produce such gas mass distributions across phases at all redshifts presented here. 

We compare the radial velocities with the escape velocity at \rgal. To understand whether the outflowing gas is gravitationally bound, we compare these velocities with the local escape velocity shown in the four panels with dashed lines. While a radial velocity higher than $v_{\rm esc}$ is not a sufficient condition for gas to completely escape the gravitational potential of a halo, it is a useful proxy commonly used in observations and theoretical studies. The radial velocities of the hot phase (top left panel) range from $\sim 100\, \rm km\,s^{-1}$ for $z\geq 3$ to $\sim 80\, \rm km\,s^{-1}$ at lower redshifts in the no-CR simulations. For the case of CRMHD, the velocities are consistently higher, being in the range  $\sim 100-200\, \rm km\,s^{-1}$. These outflow-weighted velocities are in agreement with observations of the warm ionised gas in local dwarf galaxies by \citet{McQuinn2019GalacticGalaxies} and \citet{Marasco2023ShakenGalaxies}, while falling short of the $\sim 400-600\, \rm km\,s^{-1}$ value reported in \citet{Concas2022BeingHosts} for their sample of galaxies between $1.2<z<2.6$. However, the CRMHD simulation is the only one consistently able to reach outflow velocities above the escape velocity within the galactic region, up until $z\sim 2.5$, when its distribution becomes similar to the non-CR cases. This behaviour is in agreement with the histograms shown in Fig.~\ref{fig:PDF_flowrate_comp}. 

Outflows in the warm neutral and cold phases are rarely found in the HD and MHD runs (except during the merger taking place at $z\sim 3-2$), and only appear as rare and sharp peaks in the $v_r$ plots. By examining projections of the galactic region of \nut~at the redshifts corresponding to these peaks, we have found that for HD and MHD all cold and warm neutral outflows are linked to extended tidal features of satellites, and are not spatially related to the rest of hot and warm ionised outflowing gas. In fact, CRMHD is the only run that shows warm neutral and cold outflows across the majority of snapshots. These outflows are also spatially correlated with the hotter phases of the outflowing gas and are not dynamically linked with tidally stripped satellites (see Section~\ref{subsubsec:CR_support} and Fig.~\ref{fig:outflow_proj_z2.41} for further details). Contrarily to the hot and warm ionised phases, the warm neutral and cold gas appear to show an evolution in outflow velocity, increasing from $\sim 50\, \rm km\,s^{-1}$ at high redshift to a maximum of $\sim 100\, \rm km\,s^{-1}$ at $z\sim 2.5-2$. Similarly to the warm ionised phase, the median $v_r$ of these two phases is below $v_{\rm esc}$ (except for peaks at $\sim 2.2$ and $\sim 2.5\,\rm Gyr$ in the cold phase). Overall, we find that these colder outflows struggle to reach the escape velocities at \rgal, even in the CRMHD run and despite its shallower gravitational potential.

\subsubsection{Cosmic ray-driven re-acceleration of outflows in the CGM}\label{subsubsec:CR_support}
\begin{figure*}
	\includegraphics[width=\textwidth]{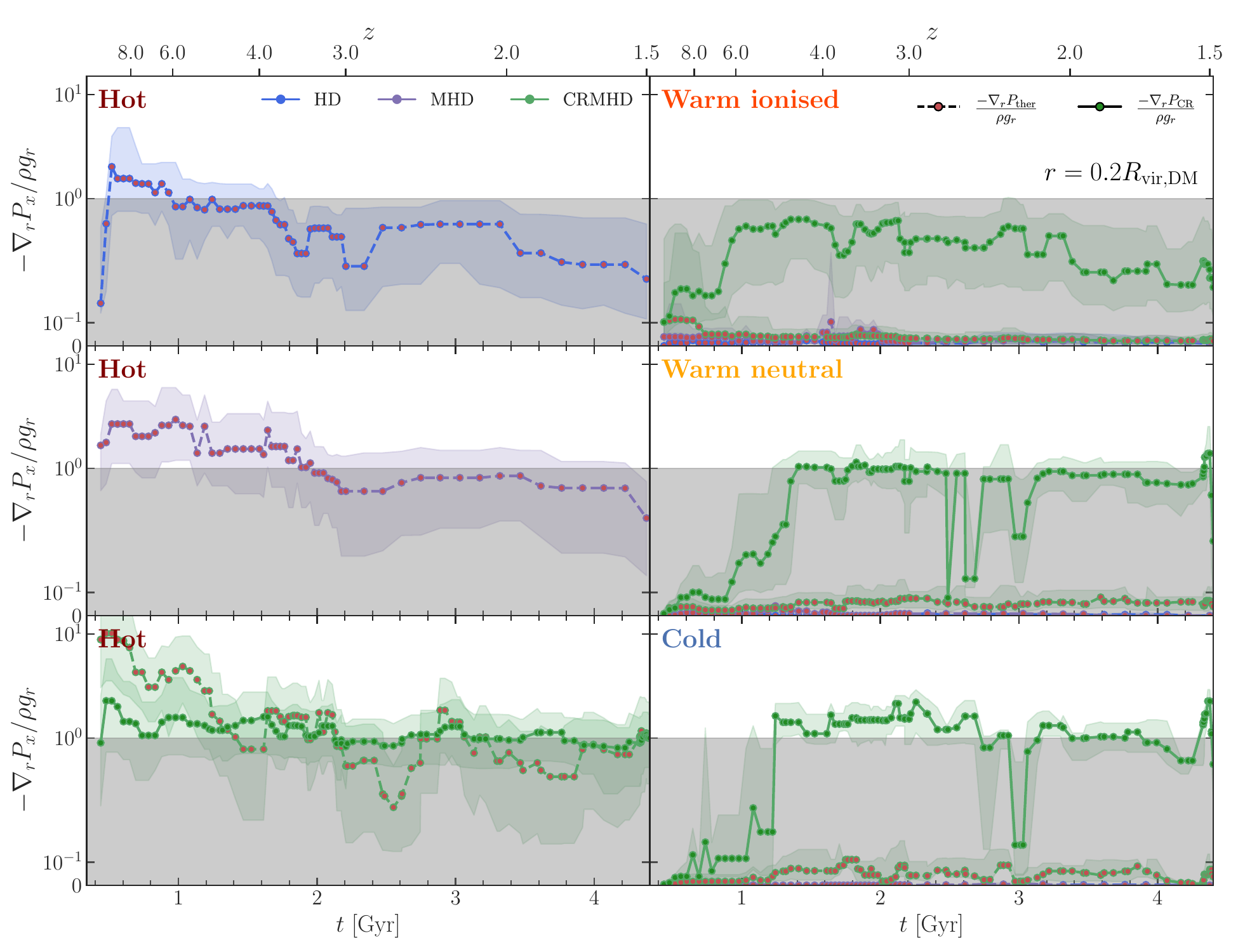}
    \caption{Pressure gradient support against gravity for outflowing gas separated into different temperature phases for HD (blue), MHD (purple) and CRMHD (green) simulations. The dividing line between the shaded and unshaded region indicates marginal support. In the shaded region, gravity dominates the dynamics, while in the unshaded region, gas is accelerated outwards by the pressure gradients. Outflowing gas is measured in a thin shell at \rgal~and divided into the phases described in Section~\ref{subsubsec:outflows} according to temperature. Shaded, coloured bands represent the second and fourth quartiles of the underlying distribution. The left column corresponds to the hot gas, with each row corresponding to the HD (top), MHD (middle), and CRMHD (bottom) simulations, respectively. While below $z\sim 3$ the thermal pressure struggles to accelerate gas in the HD and MHD simulations, CR pressure consistently supports and accelerates the hot gas across all cosmic time. Only CR pressure can mildly support the warm ionised phase between $3 < z < 6$, but it supports and accelerates warm neutral and cold gas for $z\leq 5$.}
    \label{fig:outflow_support_20rvir}
\end{figure*}

The CRMHD simulation displays considerably higher outflow rates at \rgal\ than the HD and MHD runs. However, the previous analysis has shown that gas usually lacks the kinetic energy required to fully escape the gravitational potential of the galaxy. In addition, outflowing gas is further affected by the interaction with the CGM. Therefore, in this section, we investigate how such an interaction further impacts the gas dynamics at \rgal\ in Fig.~\ref{fig:outflow_support_20rvir}. This is carried out by computing the mass-flow-rate-weighted median of the radial pressure support for the outflowing gas in the thin shell at \rgal. As per the hydrostatic equilibrium equation, the negative pressure gradient of the gas needs to balance the inward pull of gravity for a parcel of gas not to inflow. The gravitationally-dominated regime of which this equilibrium situation is a limiting case, is indicated by the grey band below a value of 1. For a pressured fluid $x$, this corresponds to $- \hat{r}\cdot \vec{\nabla} P_x = \hat{r}\cdot(\rho  \vec{g})$, where $\hat{r}$ is the radial unit vector, $\vec g$ is the local gravitational acceleration and $x$ is the fluid pressure component (either thermal or CR) considered. Cells with radial support below unity experience an overall deceleration whereas those with a value $> 1$ experience an outward acceleration. We represent the contributions from thermal pressure, $P_{\rm ther}$, as solid circle symbols coloured in red linked by dashed lines, and CR pressure, $P_{\rm CR}$, as green circles connected by solid lines. 

We begin by comparing the support for the hot phase, where we separate for clarity the HD (top left), MHD (middle left) and CRMHD (bottom left) simulations. We found that the hot and warm ionised phases show a large variance for these quantities, so we include the second and fourth quartiles of the underlying distribution as a shaded band in order to exemplify this. The three runs display a large amount of thermal pressure support against gravity for $z\geq 3$.  At lower redshifts this pressure support ratio rarely exceeds unity (although it is more common for the MHD case), and has a more bursty behaviour. This change in the support of hot outflowing gas in all simulations is attributed to the growth of the virial shock at lower redshifts, as the cooling rate becomes smaller than the post-shock compression rate \citep{Dekel2006} and cold accretion becomes less important \citep{Powell2011}. A hotter CGM means a larger thermal pressure confining gas to the galactic disk, and buoyancy forces present at high redshift becomes less important. Although, this behaviour is observed for all runs, the CRMHD run features a larger frequency of radial support peaks exceeding the force balance threshold, despite the higher density of the CGM in the CRMHD simulation compared with the non-CR simulations (see Section~\ref{subsec:global_effects}). We attribute this to a colder CGM with a lower fractional contribution from thermal support due to the additional support of CR pressure (see Section~\ref{subsec:global_effects}). We do not include the contribution of magnetic support, as we find it to be negligible for outflows at all times for the MHD and CRMHD simulations\footnote{Nevertheless, this should be considered also in the context of magnetic field injection by SN remnants \citep[e.g.][]{Martin-Alvarez2021}, which we do not explore in this study.}. Thermal pressure dominates the hot phase in CRMHD down to $z\sim 3$, with both CRs and thermal pressures separately capable of supporting a considerable fraction of the outflowing gas against gravity. The CR pressure support of the hot component evolves only mildly with redshift, eventually dominating for $z < 3$. This change in behaviour coincides with the drop in hot gas radial velocity after $z\sim 3$ shown for CRMHD in Fig.~\ref{fig:outflow_phases_20rvir}. While hot fast outflows in the HD and MHD simulations find it harder to travel through a the CGM at lower redshifts, in the CRMHD the additional support of CRs injected by SNe to the wind bubble further accelerates them when their thermal buoyancy stalls.

\begin{figure*}
    \includegraphics[width=\textwidth]{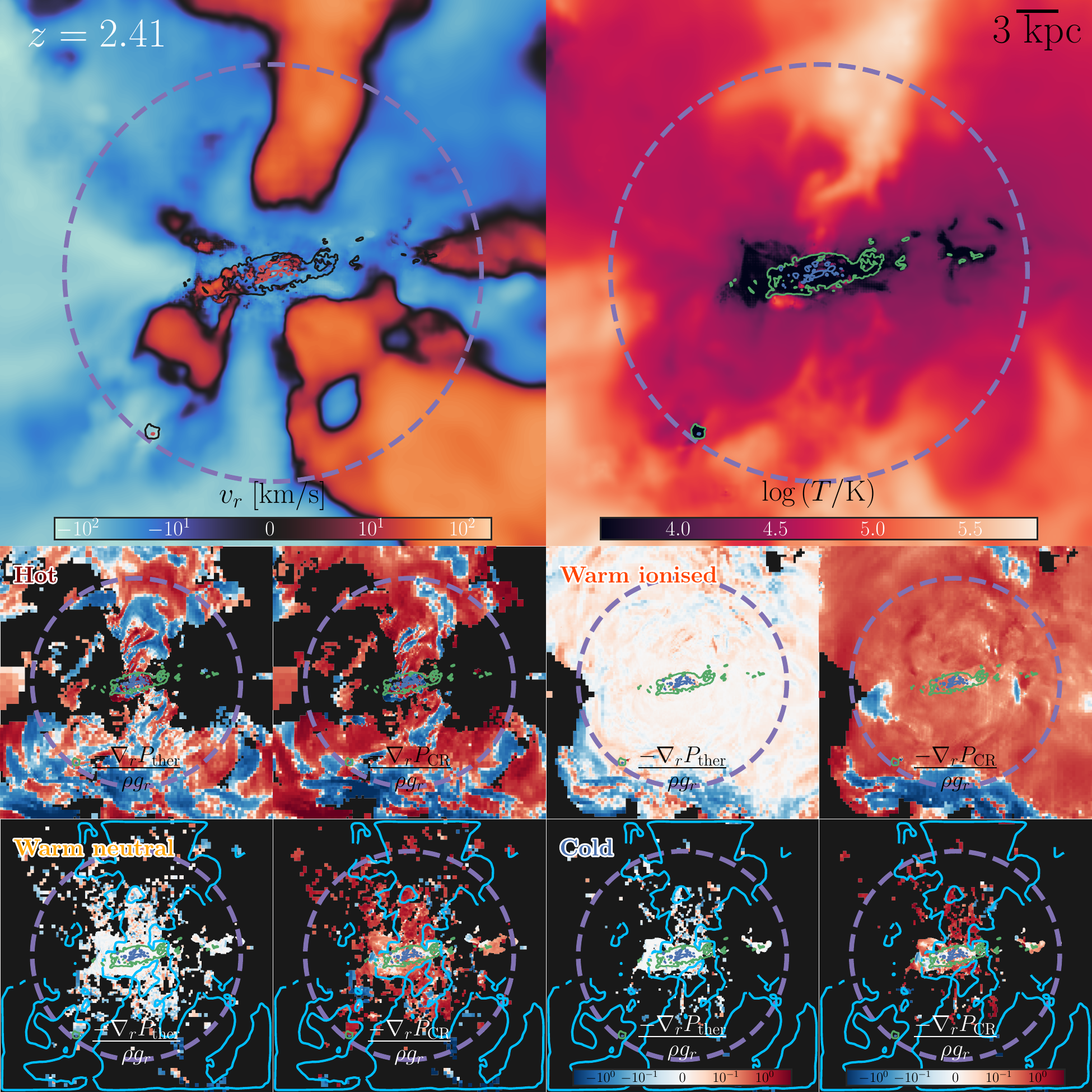}
    \caption{Edge-on view of \nut~in the CRMHD run during a strong feedback event at $z=2.41$. \textbf{Top panels:} density-weighted projections of a $20\,\rm kpc$ on a side cube centred on \nut~showing the gas radial velocity (left panel) and the gas temperature (right panel). In the left panel, inflowing gas is shown in blue and outflowing gas in red. The temperature map indicates that considerable hot outflows are ejected perpendicular to the cold disk, while warm gas at $\sim 10^{4.5}\,\rm K$ surrounds the galaxy. Black (green) and red (blue) contours are overlaid on the radial velocity (and temperature) projection to show gas density contours at $10^{-24}\,\rm g\, cm^{-3}$ and $10^{-22}\,\rm g\, cm^{-3}$, respectively. The purple dashed circle shows the \rgal\ shell used to measure gas flows close to the galactic region. \textbf{Bottom panels:} the same projected region now separated into four outflow gas phases (see text for detail). Each phase is plotted in two panels, corresponding to thermal (i.e. $-\nabla_r P_{\rm ther}/(\rho g_r)$) and CR (i.e. $-\nabla_r P_{\rm CR}/(\rho g_r)$) support of the gas against gravity, with red (outwards acceleration) or blue (inwards acceleration) colours indicating the dominant pressure gradient. From top left to bottom right, the temperature phases displayed are: hot (two top leftmost panels), warm ionised (two top rightmost panels), warm neutral (two bottom leftmost panels), and cold (two bottom rightmost panels). To help visual analysis, we include the density contours (red and blue) of the temperature panel in these panels. The warm neutral and cold panels also include contour plots showing the regions dominated by the hot (blue contours) outflows. Hot outflows are clearly biconical with large gradients of both thermal and CR pressure. Both feature alternative positive and negative gradients, a tell-tale signature of consecutive outflow waves. The warm ionised gas is the most volume filling phase, with a much more isotropic structure around \nut, and dominated by a uniform $-\nabla_r P_{\rm CR}/(\rho g_r)$ profile. Warm neutral and cold outflows present a more clumpy and filamentary structure which quickly fades off at larger radii.}
    \label{fig:outflow_proj_z2.41}
\end{figure*}

We compare in the right panels of Fig.~\ref{fig:outflow_support_20rvir} the pressure support against gravity in the warm ionised, warm neutral and cold phases, now for the three simulations combined. As warm and cold outflows are scarce in the HD and MHD runs, we limit our analysis for these phases to the CRMHD run. These confirm that the thermal pressure support (as well as the magnetic, which we do not include) in the HD and MHD runs is negligible. This further supports the idea that for the runs without CRs, outflowing gas colder than $10^5$ K does not experience further acceleration at \rgal. Due to outflow velocities lower than $v_{\rm esc}$, outflows likely rain back down onto the galaxy as a galactic fountain. The CRMHD run features warm ionised outflows that are marginally supported against gravity, following an overall trend similar to that of the hot gas, but now with CR pressure dominating over thermal pressure at all times. In fact, CR pressure support dominates over thermal pressure for the three phases in the right column of Fig.~\ref{fig:outflow_support_20rvir}. The warm neutral and the cold components of the outflow show large accelerations due to CR pressure, exceeding by up to 7 times the gravitational acceleration. These phases display a large variability, also found in the analysis of $v_{\rm r}$ (see Fig.~\ref{fig:outflow_phases_20rvir}). This can be attributed to a combination of outflow `burstiness' and a lower volume-filling of these phases. Therefore, whenever outflows entraining warm neutral and cold gas are present, they are usually subject to large CR pressure gradients that not only balance gravity but provide additional acceleration. This means that, although the outflow velocities of these two phases at these radii are not enough to escape the halo, CR pressure gradients can drastically change their dynamics as winds escape from the galaxy.

This section has described how CRs affect outflows, providing further insight on the consistently higher outflow rates that the CRMHD run exhibits when compared to the HD and MHD runs. Here we perform a more detailed analysis of a particular feedback event in the CRMHD simulation to understand how CRs are locally driving the observed differences. In Fig.~\ref{fig:outflow_proj_z2.41} we show edge-on density-weighted projections of \nut~during a feedback event at $z=2.41$ ($t\sim 2.9\,\rm Gyr$). This time was selected because of the presence of a corresponding peak in the outflow rate (i.e. $\sim 30 \,\rm \msun\,yr^{-1}$, see Fig.~\ref{fig:instantaneous_outflow_20rvir}) and the CR-supported warm neutral and cold outflows. The top panels show the radial velocity (left) and gas temperature (right) maps for a region $20$~kpc on a side, encompassing the \rgal\ thin shell (shown as a dashed purple circle). We overlay on top of the left (right) map contours for the gas density showing \qty[print-unity-mantissa = false]{e-24}{\gram\,\centi\metre^{-3}} in black (red) and at \qty[print-unity-mantissa = false]{e-22}{\gram\,\centi\metre^{-3}} in red (blue). These panels uphold the canonical understanding of wind launching in disk galaxies: at the centre lies a cold and dense disk from which high velocity, hot outflows carve biconical perpendicular channels through the warm ionised CGM. This particular outflow is preceded by previous plumes of hot gas visible at larger radii. There are also regions of outflowing gas that do not have a significantly higher temperature than the surrounding CGM (see outflowing region in the top left corner of the $v_{\rm r}$ map). 

In keeping with our previous analysis, we separate the outflowing gas into distinct temperature phases in the 8 lower panels. For each of the phases, we plot radial support against gravity, i.e. $- \nabla_r P_x = \rho  g_r$, for thermal ($P_{\rm ther}$) and CR ($P_{\rm CR}$) pressures. To help the analysis, we keep the density contours (green and blue lines) of the temperature projection in these panels. We first examine the hot outflows (top left panels), which display in both thermal and CR pressure support maps a clear biconical morphology with opening angle of $\sim 30^{\circ}$. Dark red and blue colours represent high pressure acceleration anti-parallel and parallel to the gravitational acceleration, respectively. The intertwined structure of positive and negative gradients in both thermal and CR pressure is attributed to the succession of multiple outflow bubbles, ejected from the galactic disk at different times. As such, the front of a previous bubble appears as a positive gradient in the radial direction. These features are also present in the pressure support maps of the HD and MHD simulations (not shown). This reaffirms the selection of the positive support part of these outflows  (see Fig.~\ref{fig:outflow_support_20rvir}) as a method to identify galactic winds. In agreement with the peak observed at $t\sim 2.9\,\rm Gyr$ in Fig.~\ref{fig:outflow_support_20rvir}, the thermal and CR supports are much larger than unity, both reaching similar magnitudes. Nevertheless, large positive or negative thermal and CR pressure gradients do not always spatially coincide. We surmise that such offsets could be due to CR streaming, as on $\sim\,\rm kpc$ scales it can be an important driver of CR transport and displace CRs with respect to the gas, while maintaining the CR pressure gradient \citep[e.g.][]{Dubois2019}.

The warm ionised phase of the gas is the dominant volume-filling phase of the outflow, and has a more spherically isotropic morphology. This phase does not display strong alternating features from consecutive wind bubbles, but rather a smooth pressure gradient. This pressure support extends to large radii and is dominated by CR pressure. In fact, the warm ionised, warm neutral and cold phases exhibit the same support morphology, which suggests that they are all sourced by a common underlying CR pressurised halo that surrounds the galaxy. However, based on the mild support of CR pressure on the warm ionised phase in Fig.~\ref{fig:outflow_support_20rvir}, we claim that this phase will probably rain down as a galactic fountain given the low average radial velocity (see top right panel of Fig.~\ref{fig:outflow_phases_20rvir}). The four bottom panels examine the warm neutral (left) and cold (right) outflows. Gas in these phases is distributed in small structures that resemble cooling filaments. Blue contours are included to indicate the limits of the hot outflow cones. These colder phases appear to also be moving perpendicularly to the disk, experiencing a uniform CR pressure gradient that dominates their gravitational acceleration. This indicates that these cold regions of gas are still efficiently accelerated by the surrounding hotter phase, and hence entrained in the wind. The radial distribution of these outflowing phases reveals that they decrease in fraction of volume occupied as a function of radial distance, thus allowing to better showcase their clumpy nature. These small outflowing gas structures do traverse the \rgal\ shell we employ to measure outflows (illustrated by a dashed purple circle) in a timescale shorter than our snapshot time resolution, naturally resulting in the fluctuating behaviour of warm neutral and cold phases present in Figs.~\ref{fig:outflow_phases_20rvir} and~\ref{fig:outflow_support_20rvir}.

\section{Conclusions}\label{sec:conclusions}

In this work, we have performed a detailed study of the effects of CR feedback on a Milky Way-like galaxy using a suite of new zoom-in cosmological simulations. The main galaxy in our zoom region, \nut, was simulated down to $z=1.5$ with the \ramses\ code. Using adaptive mesh refinement, we simulated the multi-phase ISM with a spatial resolution of $\sim 23\,\rm pc$ per full cell size. Our hydrodynamical simulation (HD) includes a model for a resolved ISM, with a magneto-thermo-turbulent (MTT) star formation prescription and a 'mechanical' SN feedback of a canonical strength. In our magnetohydrodynamic simulation (MHD) we additionally followed the magnetic fields with the constrained transport method which are amplified from a primordial seed of $3\times 10^{-12}\,\rm G$. Our cosmic ray MHD simulation (CRMHD) further models CRs injected by supernova (SN) events with an efficiency of 10\%. In this first work in a series of papers analysing the role of CRs in the \nut~suite, we focused on the impact of our full CR physics which includes advection, losses, anisotropic diffusion and streaming on the ISM and the launching of outflows. Our main results can be summarised as follows.
\begin{enumerate}
    \item Including CRs significantly reduces the resulting stellar mass of \nut~ compared to the no-CR models, yielding good agreement with abundance matching models \citep[e.g.][]{Behroozi2013The0-8,Moster2013GalacticHaloes} without the need for a calibrated boost to the SN energy deposition or calibrated galactic winds. This reduction in stellar mass is significant and it reaches approximately an order of magnitude at high redshift $z\gtrsim 3$ before tapering to a factor $\sim 4$ at $z\sim 1.5$ (Section~\ref{subsec:global_effects} and Fig.~\ref{fig:smf}). 
    \item CRs pressure gradients support cold, low density gas against collapse, making the warm phase significantly more dominant in the ISM (Section~\ref{subsec:sf_efficiency}). CRs also deplete an important fraction of the mass locked in the cold, dense phase (i.e. $\rho\gtrsim 10^{-21}\,\rm g\,cm^{-3}$ and $T < 10^3\,\rm K$) present in non-CRs simulations, and hence significantly decrease the amount of gas efficiently forming stars. In fact, galactic gas is rarely found below $10^{-27}\,\rm g\,cm^{-3}$ and the ISM has a smoother gas distribution within the disk. 
    \item As a consequence of the thermodynamical changes to the structure of the ISM and a CR-aided expansion of SN bubbles \citep[e.g.][]{RodriguezMontero2022MomentumRays}, the environments where SN events take place are modified. In the CRMHD simulation, SN explosion sites have a bi-modal distribution that both peaks at and reaches considerably lower densities than in the HD and MHD runs (see Fig.~\ref{fig:sne_pdfs}). This reduces the amount of SN energy lost due to cooling, increases the efficiency of this feedback mechanism and drives stronger outflows.
    \item The mass outflow rates in the CRMHD simulation are significantly larger at all cosmic times with respect to the no-CR runs, specifically with a higher mass loading factor $\eta$ especially at high redshift. In contrast to the non-CR simulations where the dominant phase of outflowing gas mass is hot (i.e. $T > 10^5\,\rm K$), CRMHD outflows are dominated in mass by the warm ionised phase (i.e. $9000 < T < 10^5\,\rm K$) and have a significant contribution from the cold phase (see Section~\ref{subsubsec:outflows} and~\ref{subsubsec:phases}). Furthermore, hot outflows are biconical in structure whereas the colder phases embedded in the hot outflow are filamentary and clumpy. This is a direct prediction for upcoming JWST observations of multi-phase galactic outflows expected soon.
    \item Outflows in the CRMHD simulation experience further acceleration due to the CR pressure gradient, with a secondary contribution from the thermal pressure. Outflows in the warm and cold phases have a significant support from the CRs, while the thermal pressure support remains negligible. Therefore, outflows experience further acceleration within the CGM of our simulated galaxy (see Section~\ref{subsubsec:CR_support} and Fig.~\ref{fig:outflow_proj_z2.41}).
\end{enumerate}

We find CRs to have a large impact across various properties of the simulated galaxy and its ISM. While we include multiple important CR physical components in our simulations, the specific impact of each is not fully explored. A review of these roles is left to the next paper in this series, which will analyse the effects of this CR feedback on the thermodynamic state of the CGM outside the \nut~ galaxy (Rodr\'iguez Montero et al. \textit{in prep.}). 

While simulations, such as the ones presented here, are revealing CRs to be a central agent in galaxy formation, our CR models are far from being complete and our results suffer from multiple caveats, both physical and numerical. For instance, we still lack a self-consistent coupling of CR transport with the local thermodynamic state of the gas \citep[e.g.][]{Armillotta2022Cosmic-RayEnvironments}. Furthermore, CR physics may well be affected by more realistic and complete gas cooling prescriptions \citep{Nunez-Castineyra2022Cosmic-rayLuminosities,Katz2022PRISM:Galaxies} and different refinement criteria in the CGM \citep{Bennett2020ResolvingMedium, Rey2023BoostingResolution}. Milky Way-like halos at high redshift can have their star formation strongly altered by other forms of feedback such as that from stellar winds \citep{Agertz2021VintergatanGalaxy}, AGN \citep{Koudmani2022TwoSimulations}, or radiation from young stars\citep{Rosdahl2015ARAMSES-RT,Hopkins2020a,Emerick2018StellarGalaxies}. Their self-consistent interaction with CR feedback remains a topic for further investigation \citep[e.g.][]{Farcy2022Radiation-MagnetoHydrodynamicsGalaxies,Martin-Alvarez2022TheGalaxies}. 

Despite these caveats, our work highlights the fundamental role CRs likely play in the formation and evolution of Milky Way-like galaxies. In particular, they seem to be a central contributor to the feedback loop regulating star formation and driving galactic outflows, providing a promising avenue to form galaxies with more realistic physical properties without having to resort to ad-hoc feedback model calibrations.

\section*{Acknowledgements}

FRM is supported by the Wolfson Harrison UK Research Council Physics Scholarship. The authors thank Romain Teyssier for making the \ramses~code publicly available. This project has received funding from the European Research Council (ERC) under the European Union’s Horizon 2020 research and innovation programme (grant agreement No 693024). Simulations were performed using the DiRAC Data Intensive service at Leicester, operated by the University of Leicester IT Services, which forms part of the STFC DiRAC HPC Facility (\href{www.dirac.ac.uk}{www.dirac.ac.uk}). The equipment was funded by BEIS capital funding via STFC capital grants ST/K000373/1 and ST/R002363/1 and STFC DiRAC Operations grant ST/R001014/1. DiRAC is part of the National e-Infrastructure. Analysis of the simulations took place on the Oxford Astrophysics cluster `Glamdring', that we warmly thank J.~Patterson for running smoothly.

\section*{Data Availability}

 The simulation data underlying this article will be shared on reasonable request to the corresponding author.



\bibliographystyle{mnras}
\bibliography{references} 



\bsp	
\label{lastpage}
\end{document}